\documentclass[useAMS, usenatbib]{mn2e}
\usepackage{times}
\usepackage[final]{graphicx}
\usepackage{amssymb} 
\usepackage{breqn}
\usepackage{subfig}
\newcommand{\sech}{\, \mathrm{sech} }
\newcommand{\dz}{\, \mathrm{dz} \, }

\title[Structure formation in gas-rich galactic discs]{Structure formation in gas-rich galactic discs with finite thickness: \\ from discs to rings}
\author[M. Behrendt et al.]{M. ~Behrendt$^{1,2}$\thanks{E-mail: mabe@mpe.mpg.de},
		A. ~Burkert$^{1,2}$\thanks{Max Planck Fellow},
		M. ~Schartmann$^{1,2}$ \\
$^1$Max Planck Institute for extraterrestrial Physics, PO box 1312, Giessenbachstra${\ss}$e, D-85741 Garching, Germany\\
$^2$University Observatory Munich, Scheinerstra\ss e 1, D-81679 Munich, Germany}
\begin{document} 

\date{Accepted 2015 January 5. Received 2015 January 5; in original form 2014 July 20}

\pagerange{\pageref{firstpage}--\pageref{lastpage}} \pubyear{2015}

\maketitle

\label{firstpage}

\begin{abstract}
Gravitational instabilities play an important role in structure formation of gas-rich high-redshift disc galaxies. In this paper, we revisit the axisymmetric perturbation theory and the resulting growth of structure by taking the realistic thickness of the disc into account. In the unstable regime, which corresponds for thick discs to a Toomre parameter below the critical value $Q_{\mathrm{0,crit}} = 0.696$, we find a fastest growing perturbation wavelength that is always a factor 1.93 times larger than in the classical razor-thin disc approximation. This result is independent of the adopted disc scaleheight and by this independent of temperature and surface density. In order to test the analytical theory, we compare it with a high-resolution hydrodynamical simulation of an isothermal gravitationally unstable gas disc with the typical vertical $\sech^2$ density profile and study its break up into rings that subsequently fragment into dense clumps. In the first phase, rings form, that organize themselves discretely, with distances corresponding to the local fastest growing perturbation wavelength. We find that the disc scaleheight has to be resolved initially with five or more grid cells in order to guarantee proper growth of the ring structures, which follow the analytical prediction. These rings later on contract to a thin and dense line, while at the same time accreting more gas from the inter-ring region. It is these dense, circular filaments, that subsequently fragment into a large number of clumps. Contrary to what is typically assumed, the clump sizes are therefore not directly determined by the fastest growing wavelength.
\end{abstract}


\begin{figure*}
\centering
\subfloat[  \label{fig:reductionfactor}]
  {\includegraphics[width=.48\linewidth]{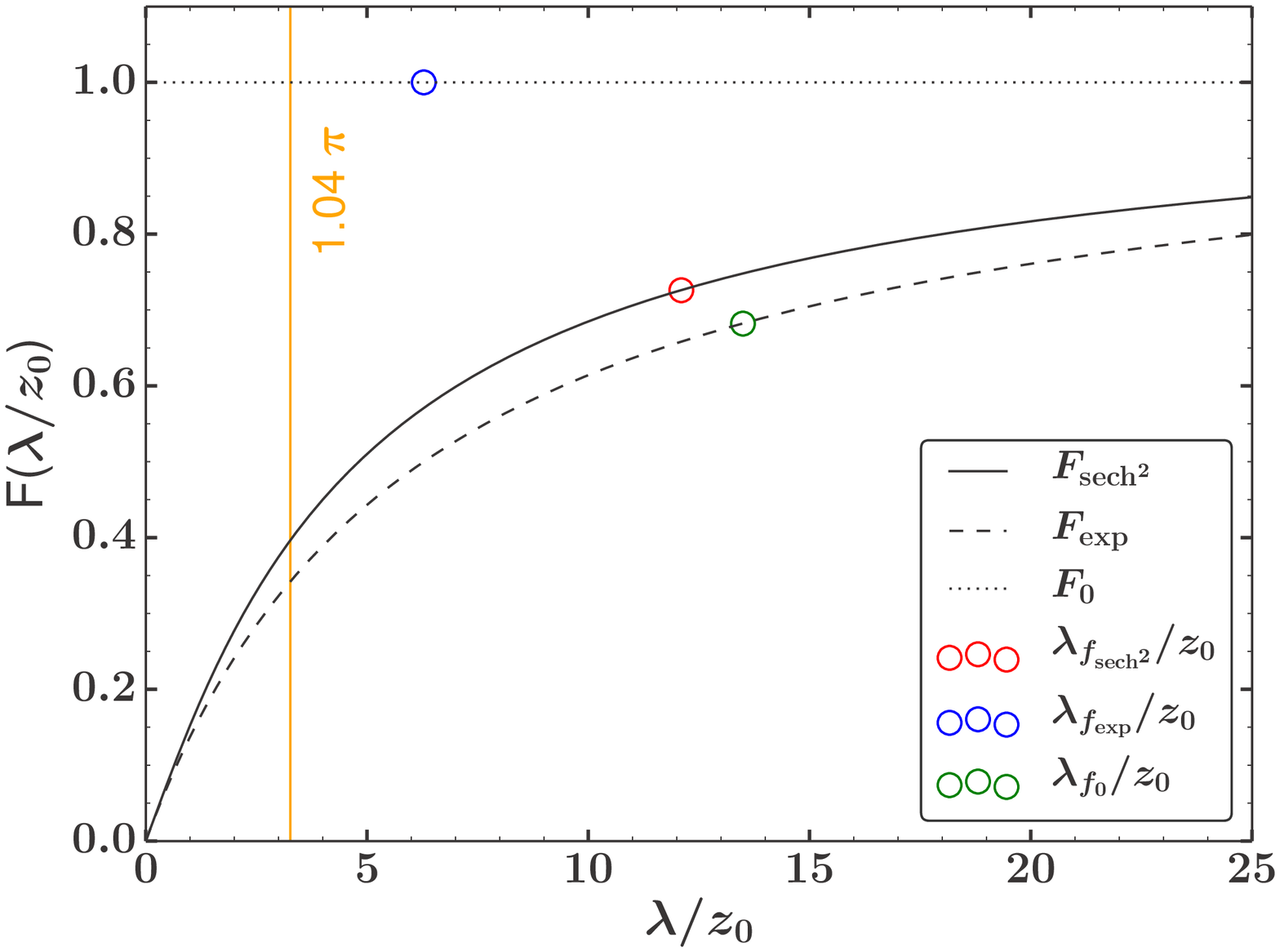}}\hfill
\subfloat[ \label{fig:reductionfactor_error}]
  {\includegraphics[width=.48\linewidth]{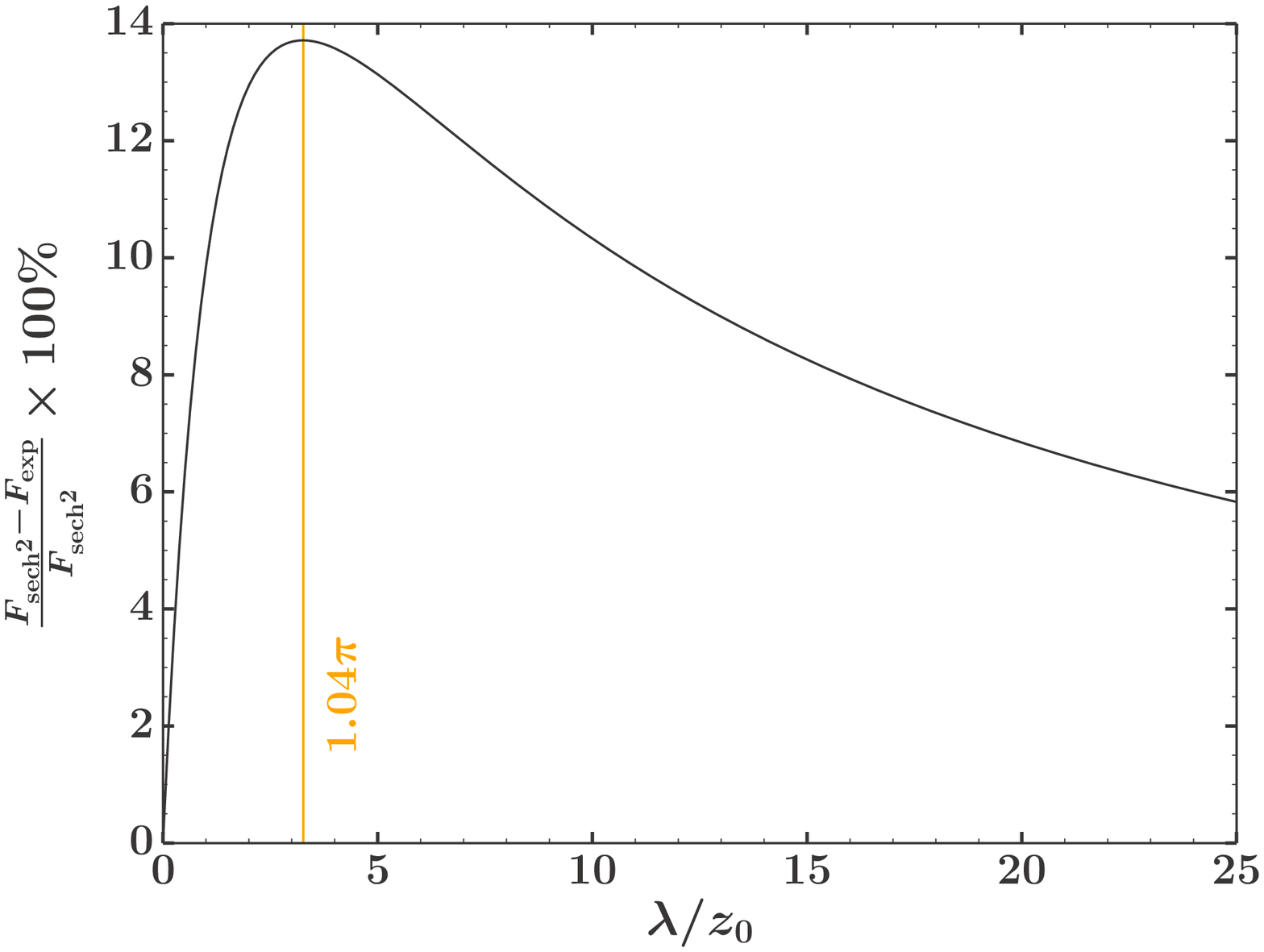}}\hfill
\caption{(a) General dependence of the reduction factor $F( \lambda / z_0)$. The numerical integration of $F_{\mathrm{sech^2}}$ is represented by the solid line (equation\ref{eq:F_sech2_rephrased}), the approximation $F_{\mathrm{exp}}$ by the dashed line (equation \ref{eq:F_exp_rephrased}) and with no reduction for an infinitesimally thin disc by the dotted line. The coloured open circles mark the particular reduction for the constant relations between the scale-height and the fastest growing wavelength derived for the different vertical density distributions (Sections \ref{subsec:Approximated fastest growing perturbation wavelength} and \ref{subsec:Exact fastest growing perturbation wavelength}). (b) Per cent deviation between the reduction factor for the $\sech^2$ profile and the approximation with the exponential profile, with the maximum of $13.7 \%$ at $1.04 \upi$. }
\end{figure*}

\section{Introduction}
\label{sec:introduction}
Observations of massive galaxies at high redshift have revealed rotating discs \citep{2009ApJ...706.1364F}  with large gas fractions \citep{2010ApJ...713..686D} and high velocity dispersion. Their morphology is irregular and often dominated by a few kpc-sized clumps \citep{2011ApJ...733..101G}, seen in the optical \citep{2007ApJ...658..763E}, and also in ionized \citep{2008ApJ...687...59G} and molecular gas \citep{2013ApJ...768...74T}. In the Cosmic Assembly Near-IR Deep Extragalactic Legacy Survey \citep[CANDELS;][]{2011ApJS..197...35G,2011ApJS..197...36K} over 250 000 distant galaxies from $z = 8$ to $1.5$ are documented, where more than half of the star-forming galaxies are clumpy \citep[][and references therein]{2014ApJ...780...57B}. The main origin of bound clumps is expected to be a result of in situ gravitational disc fragmentation, as demonstrated by a large number of numerical simulations of self-gravitating gas-rich discs, either in a cosmological context \citep{Agertz:2009wd,Dekel:2009bn,Ceverino:2010eh,2012MNRAS.420.3490C} or in an isolated box \citep{2004A&A...413..547I,2004ApJ...611...20I,2007ApJ...670..237B,2012MNRAS.420.3490C}. In general, local gravity has to overcome the stabilizing effects of thermal or turbulent pressure and differential rotation. This is quantified by the parameter $Q_0$ of \citet{1964ApJ...139.1217T}. Below a critical value ($Q_0<1$), axisymmetric instabilities (rings in an axisymmetric disc) can form, contract and finally break up into several clumps \citep{Dekel:2009bn}, faster than spiral patterns (non-axisymmetric modes) can form. Furthermore, to finally form strongly bound clumps that later on collapse to stars, sufficient cooling is required to ensure contraction to higher densities within short enough time-scales \citep{Gammie:2001hv,Dekel:2009bn}. \\ 
A more detailed insight can be achieved from linear perturbation theory \citep{Lin:1964cx}. It can describe the properties of structure formation, with respect to their sizes, masses and time-scales. Understanding these processes in detail is important in order to get a better view of the previous history and subsequent evolution of a disc. However, in these considerations, a razor-thin disc was assumed, characterized by its surface density $\Sigma$. Already \citet{1964ApJ...139.1217T} pointed out that a disc with finite thickness leads to a reduction of the gravitational force in the mid-plane, where structure formation takes place. Therefore, the disc remains stable even below a $Q_0$-value of one and a revised critical value has to be determined with $Q_0 < Q_{\mathrm{crit}} < 1$, describing the unstable regime. \\
\\
A self-gravitating gas disc with its isothermal vertical structure in hydrostatic equilibrium is given by the $\sech^2$ density profile \citep{1942ApJ....95..329S}. Since there is no analytical solution for the reduction factor of the potential in the mid-plane for this profile, a good approximation is achieved by using an exponential function instead \citep[e.g.][]{1987ApJ...312..626E}, for which the reduction factor deviates from the one of the $\sech^2$ profile with a maximum error of $14\%$ \citep{Kim:2002gl, 2006ApJ...649L..13K,Kim:2007ek}. The critical value for the instability parameter was thus determined to $Q_{\mathrm{crit}} \simeq 0.647$ for an exponential profile approximation by \citet{Kim:2007ek} and for the $\sech^2$ profile to $Q_{\mathrm{crit}} \simeq 0.693$ in \citet{Wang:2010wr}. Moreover, the reduced self-gravity also affects the sizes of the growing structures. \citet{2012MNRAS.422..600G} estimated for an exponential vertical distribution a factor of $\sim 2$ larger wavelengths for the fastest growing perturbations. These implications on stability and structure properties demonstrate the importance of considering the thickness effects.\\
\\
In this paper, we go beyond previous work by investigating in detail how axisymmetric instabilities grow in an idealized gas disc simulation. Section \ref{sec:Analytical model} introduces the linear stability analysis for thick discs. From this we deduce the properties of the growing ring structures with respect to their sizes, masses and formation time-scales. Finally, we employ an idealized hydrodynamical simulation (Section \ref{sec:Numerical modelling}) and compare the initial structure formation with the linear stability analysis (Section \ref{sec:Results}).

\section{Analytical model}
\label{sec:Analytical model}
We summarize the available literature on how perturbation theory has to be modified when moving from an infinitesimally thin disc to a more realistic disc with finite thickness. In general, this transition is expressed by a wavelength and scale-height dependent reduction factor of the perturbed potential. Based on the resulting modification of the dispersion relation, we explicitly present the analytical derivation of the fastest growing wavelength for a vertically exponential density distribution. The reduction factor is independent of the scale-height when the disc is in hydrostatic equilibrium. The fastest growing wavelength for the vertical $\sech^2$ density profile is numerically approximated. At this wavelength, which is the dominant growing perturbation in a disc, the reduction factor is a constant and allows us to simplify the deduction of the physical properties of the growing ring structures and the local instability parameter.

\subsection{Reduction factor}
\label{subsec:Reduction factor}
In classical linear perturbation theory, matter is assumed to be concentrated in a razor-thin layer with a surface density $\Sigma$. This overestimates the potential as in real discs matter is vertically arranged with the density having its maximum in the mid-plane where the initial structures form first. This softens the potential and requires a correction factor $F$ in addition to modify the classical theory.\\
Density perturbations correlate with variations in the gravitational field and are usually specified by the perturbed local gravitational potential $\Phi_{\mathrm{0}}$ of an infinitesimal thin disc, which is given by the solution of the Poisson equation in the stationary form (see Appendix \ref{App:Derivation of the reduction factor of the potential due to the disc thicknes}). The total potential at the mid-plane, $z=0$, of a disc with finite thickness can then be expressed by
\begin{equation}
\label{eq:phi_tot_reduction_factor}
\Phi_{\mathrm{tot}}(\lambda, x, z=0) = \Phi_0(\lambda, x, z=0) \; F(\lambda, z=0), 
\end{equation}
with the wavelength $\lambda \geq 0$ and $x = R - R_{\mathrm{0}}$ the position near a given radius $R_{\mathrm{0}}$. The reduction factor $F(\lambda)$ is generally between two limits: $\lim_{\lambda \to \infty} F(\lambda) = 1$  and $\lim_{\lambda \to 0} F(\lambda) = 0$. Therefore, for longer wavelengths $\lambda$, $\Phi_{\mathrm{tot}}$ (equation \ref{eq:phi_tot_reduction_factor}) behaves like the potential of a razor-thin disc and for shorter $\lambda$ the potential is reduced (Fig. \ref{fig:reductionfactor}), e.g. \citet{ Kim:2002gl, Wang:2010wr}. The transition between the regimes depends on the vertical density distribution. \\
The vertical structure of a self-gravitating and isothermal sheet in  hydrostatic equilibrium is described by a $\sech^{2}(z/z_{0})$ profile \citep{1942ApJ....95..329S}, with the sound speed $c_{\mathrm{s}}$ and the scale-height
\begin{equation}
\label{eq:scale-height_} 
z_{0} = \frac{c_{\mathrm{s}}^{2}}{\upi G \Sigma}.
\end{equation}
The reduction factor is then given by (see Appendix \ref{App:Derivation of the reduction factor of the potential due to the disc thicknes})
\begin{equation}
\label{eq:integralsechhyperbolfunction} 
F_{\mathrm{sech^2}}(\lambda, z_0)  =  \int_{-\infty}^{\infty}  \mathrm{e}^{- \frac{2 \upi}{\lambda} \mid h \mid} \; \frac{\sech^{2} \left( h /z_{0} \right)}   {2 z_{0}} \; dh.
\end{equation}
This is basically the summation over all contributions to the total potential at the mid-plane, generated from infinitesimally thin layers at all vertical distances $h$. Equation (\ref{eq:integralsechhyperbolfunction}) can be integrated numerically. Alternatively, an analytical solution can be achieved by replacing the vertical $\sech^{2}$ distribution by an exponential profile, which is a good approximation  \citep{1983ApJ...267...31E, 1987ApJ...312..626E, Kim:2002gl, Wang:2010wr, 2011ApJ...737...10E}.
The reduction factor is then
\begin{equation}
\label{eq:integralexponentialfunction} 
F_{\mathrm{exp}}(\lambda, z_0)  = \int_{-\infty}^{\infty}  \mathrm{e}^{- \frac{2 \upi}{\lambda} \mid h \mid} \; \frac{\mathrm{e}^{-\mid h \mid /z_{0}} }{2 z_{0}} \; dh,
\end{equation}
and the solution
\begin{equation}
\label{eq:exp_reductionfactor}
F_{\mathrm{exp}}(\lambda, z_{0}) = \left(  1+ \frac{2 \upi }{ \lambda } \; z_{0} \right) ^{-1} .
\end{equation}
If we substitute $x = h /z_0$ in equations (\ref{eq:integralsechhyperbolfunction}) and  (\ref{eq:integralexponentialfunction}), and use the symmetry in vertical direction, the integral can be written as
\begin{equation}
\label{eq:F_sech2_rephrased}
F_{\mathrm{sech^2}}(\lambda, z_0)  =  \int_{0}^{\infty}  \mathrm{e}^{- \frac{2 \upi }{\lambda} z_0 x} \; \sech^{2}\left(x \right) \; dx,
\end{equation} 
and
\begin{equation}
\label{eq:F_exp_rephrased}
F_{\mathrm{exp}}(\lambda, z_0)  =  \int_{0}^{\infty} \mathrm{e}^{- \frac{2 \upi }{\lambda} z_0 x}  \; \exp \left(- x \right) \; dx,
\end{equation}
respectively. Both equations  (equations \ref{eq:F_sech2_rephrased} and \ref{eq:F_exp_rephrased}) show a similar dependence on $\lambda$ and $z_0$,  the comparison is illustrated in Fig. \ref{fig:reductionfactor}. In principle, for a given $z_{0}$, a smaller wavelength $\lambda$ leads to smaller $F$. Much larger wavelengths $\lambda$ on the other hand are required to reach the upper range of $F \approx 1$. The reduction factor $F_{\mathrm{exp}}$ underestimates the numerical solution of $F_{\mathrm{sech^2}}$,  with a maximum error of $13.7\%$ (Fig. \ref{fig:reductionfactor_error}), in agreement with the result found by e.g. \citet{Kim:2002gl, 2006ApJ...649L..13K} and \citet{Kim:2007ek}. This maximum error is reached however at $\lambda / z_{0} \simeq 1.04 \upi$, which differs by roughly a factor of 2 from their result.

\subsection{Modified dispersion relation}
\label{subsec:Modified dispersion relation}
In this section, we introduce the modified local dispersion relation of \citet{Lin:1964cx}  for geometrically thick gas discs, since it is important for our further calculations. It gives a relation between the angular frequency $\omega$ and density perturbations with wavelength $\lambda$ in radial direction. Here, it indicates the possibility for growing axisymmetric perturbations, depending on the local conditions. \\
In general, to get the dispersion relation, linear perturbation theory is employed on the hydrodynamical equations and the Poisson equation (Section \ref{subsec:Reduction factor}) for a self-gravitating and rotating gas disc in the razor-thin limit \citep{Binney:2011vb}. In our case, the reduction factor appears in addition and we find in the mid-plane for $\lambda > 0$ \citep[e.g.][]{Wang:2010wr}
\begin{equation}
\label{eq:modified_dispersion_relation_generalized}
\omega^{2} = \kappa^{2} - 2 \upi G \Sigma  \left(  \frac{2 \upi}{\lambda } \right)   F( \lambda, z_{0}) + c_{\mathrm{s}}^{2}  \left(\frac{2\upi}{\lambda} \right)^{2},
\end{equation}
with the epicyclic frequency $\kappa$ and the unperturbed surface density $\Sigma$. For $\omega^2 > 0$ the disc is stable against axisymmetric instabilities. If $\omega^2 < 0$, a certain range of perturbations can grow exponentially in rings. The more negative the values, the faster they grow. All parameters in equation (\ref{eq:modified_dispersion_relation_generalized}) can change with radius for the case of an exponential surface density profile, which leads to a complex interplay between the three terms. However, in general we can say: the effect of self-gravity is destabilising, whereas the differential rotation, represented by $\kappa^2 > 0$, and the thermal pressure, with $c_{\mathrm{s}}^2$, stabilize the disc. The dispersion relation (equation \ref{eq:modified_dispersion_relation_generalized}) inherits the properties of the reduction factor $F$ (see  Section \ref{subsec:Reduction factor}). For larger perturbation wavelengths, $F$ approaches unity which gives the solution of $\omega$  for a razor-thin layer. At smaller wavelengths, the thickness of the vertical density structure plays a stronger role and self-gravity is reduced in the mid-plane \citep[e.g.][]{Kim:2002gl}. With the reduction factor $F_{\mathrm{exp}}$, the relation leads, with all its dependences to the form \citep[e.g.][]{2006ApJ...647..997S,Wang:2010wr}, 
\begin{equation}
\label{eq:disp_relation_exp} 
\omega_{\mathrm{exp}}^{2} = \kappa^{2} - \frac{4 \upi^{2} G \Sigma }{\lambda + 2 \upi  z_{0} } + c_{\mathrm{s}}^{2} \left(\frac{2\upi}{\lambda} \right)^{2} .
\end{equation}
Since there is no wavelength dependence for $\kappa$, we can derive a rotation-independent form for the fastest growing perturbations in the next section.

\subsection{The fastest growing perturbation wavelength for the exponential profile}
\label{subsec:Approximated fastest growing perturbation wavelength}
The extremum of the dispersion relation as function of wavelength (equation \ref{eq:disp_relation_exp}) is determined by
\begin{equation}
\label{eq:calcfastestgrowingwavelength}
\frac{\upartial \omega_{\mathrm{exp}}^{2}} {\upartial \lambda} = 4 \upi^{2} \left( \frac{G \; \Sigma}{ (\lambda  + 2 \upi z_{0})^{2}} - \frac{2 c_{\mathrm{s}}^{2}}{\lambda^{3}}  \right)   = 0 .
\end{equation}
We solve equation (\ref{eq:calcfastestgrowingwavelength}) in Appendix \ref{App:Calculation of the fastest growing wavelength for the exp-profile approximation} analytically and get the fastest growing wavelength $\lambda_{f_{\mathrm{exp}}}$. The minimum is ensured by $\frac{\upartial^{2} \omega_{\mathrm{exp}}^{2} (\lambda_{f_{\mathrm{exp}}})} {\upartial^{2} \lambda} > 0$. It marks the maximum growth rate for an unstable disc. For a disc in hydrostatic equilibrium with given $z_0$ (equation \ref{eq:scale-height_}) the solution simplifies and we can relate it to the classical solution in the razor-thin limit with
\begin{equation}
\label{eq:fastest_growing_wavelength_thin_disk}
\lambda_{f_0}  = \frac{2 c_{\mathrm{s}}^{2}}{G \Sigma} .
\end{equation}
For the exponential profile approximation, we get
\begin{equation}
\lambda_{f_{\mathrm{exp}}} = A_{\mathrm{exp}} \; \lambda_{f_0} ,
\label{eq:fastest_growing_wavelength_exp}
\end{equation}
with the constant factor $A_{\mathrm{exp}} \simeq 2.148$ and no additional dependences. Since the form of equation (\ref{eq:fastest_growing_wavelength_thin_disk}) is similar to that of the scaleheight in equation (\ref{eq:scale-height_}), we can write
\begin{equation}
\lambda_{f_{\mathrm{exp}}} = A_{\mathrm{exp}} \; 2 \upi z_{0} \simeq 13.496 \;  z_{0},
\label{eq:fastest_growing_wavelength_z0_relation}
\end{equation}
which is similar to the approximation found by \citet{2012MNRAS.422..600G} and \citet{ 2014MNRAS.442.1230R} with $\sim 4\upi z_0$. \\
The reduction factor equation (\ref{eq:exp_reductionfactor}) then simplifies for $\lambda_{f_{\mathrm{exp}}}$ (equation \ref{eq:fastest_growing_wavelength_z0_relation}) to a constant value
\begin{equation}
\label{eq:F_exp_constant}
F_{\mathrm{exp}}(\lambda_{f_{\mathrm{exp}}})  = \left(  1 + \frac{1}{A_{\mathrm{exp}}}  \right)^{-1} \simeq 0.682,
\end{equation}
independent of the scaleheight (Fig. \ref{fig:reductionfactor}). According to the solution in equation (\ref{eq:F_exp_constant}), the potential in the mid-plane of a razor-thin disc is therefore overestimated by $46.628 \%$ (see equation \ref{eq:phi_tot_reduction_factor}).

\subsection{Fastest growing perturbation wavelength for the $\sech^2$-profile}
\label{subsec:Exact fastest growing perturbation wavelength}
We follow the steps described in Section \ref{subsec:Approximated fastest growing perturbation wavelength} and determine the minimum of the dispersion relation for the $\sech^2$ -profile by using the reduction factor $F_{\mathrm{sech^2}}(\lambda , z_0)$ with
\begin{equation}
\label{eq:sech2_dispersionrelation_peak}
\frac{\upartial \omega_{\mathrm{sech^2}}^{2}}{\upartial \lambda} = 4 \upi^{2} \left[  \frac{G \; \Sigma}{ \lambda  }  \left( \frac{F_{\mathrm{sech^2}}}{ \lambda } - \frac{\upartial F_{\mathrm{sech^2}}}{\upartial \lambda}   \right)    - \frac{2 c_{\mathrm{s}}^{2}}{\lambda^{3}}  \right]   = 0.
\end{equation}
The wavelength $\lambda$ is varied iteratively for several fixed scaleheights $z_0$ until equation (\ref{eq:sech2_dispersionrelation_peak}) reaches zero. The integral $F_{\mathrm{sech^2}}$ of equation (\ref{eq:integralsechhyperbolfunction}) and its differentiation in equation (\ref{eq:sech2_dispersionrelation_peak}) is solved numerically. We find a relation for the fastest growing perturbation wavelength that is similar to the exponential case (equation \ref{eq:fastest_growing_wavelength_exp})
\begin{equation}
\label{eq:fastest_growing_wavelength_sech2}
\lambda_{f_{\mathrm{sech^2}}} = A_{\mathrm{sech^2}} \; \lambda_{f_0},
\end{equation}
with $A_{\mathrm{sech^2}} \simeq 1.926$. Expressed by the scaleheight it is
\begin{equation}
\label{eq:fastest_growing_wavelength_sech2_z0}
\lambda_{f_{\mathrm{sech^2}}} = A_{\mathrm{sech^2}} \; 2 \upi z_0 \simeq  12.103 \;  z_0 .
\end{equation}
The error for the exponential profile approximation (equation \ref{eq:fastest_growing_wavelength_exp}) is therefore $11.527 \%$ compared to the numerical solution (equation \ref{eq:fastest_growing_wavelength_sech2}), and for the thin disc approximation $48.079 \%$ (equation \ref{eq:fastest_growing_wavelength_thin_disk}). \\
Again, we get a constant reduction factor for $\lambda_{f_{\mathrm{sech^2}}}$ for all $z_0$ with (Fig. \ref{fig:reductionfactor})
\begin{equation}
F_{\mathrm{sech^2}}(\lambda_{f_{\mathrm{sech^2}}}) \simeq 0.726,
\end{equation}
with an error of $6.061 \%$ for $F_{\mathrm{exp}}(\lambda_{f_{\mathrm{exp}}})$ compared to the integral $F_{\mathrm{sech^2}}(\lambda_{f_{\mathrm{sech^2}}})$ and $37.741 \%$ for the thin disc approximation. Due to the steepness of the exponential profile in the innermost part, matter and self-gravity, respectively, is slightly underestimated in the mid-plane compared to the $\sech^2$ distribution.

\subsection{Ring properties}
\label{subsec:Ring properties}
From the calculations above, we can derive properties of the possible growing structures if locally unstable. Due to axisymmetry, radial perturbations can grow within rings, with an overdense and an underdense part of the dominant growing wavelength. The initial radial thickness of the overdense region that will be investigated in detail in our disc simulation (see Section \ref{Results:subsec:Ring properties}), is defined as
\begin{equation}
\label{eq:half_wavelength_sech2}
L_{\mathrm{sech^2}} = \frac{\lambda_{f_{\mathrm{sech^2}}}}{2},
\end{equation}
and related to the scale-height
\begin{equation}
\label{eq:l_ring_z0}
L_{\mathrm{sech^2}} = A_{\mathrm{sech^2}} \; \upi z_0. 
\end{equation}
For the thin disc approximation, we have $L_0 = \lambda_{f_0} / 2$ and get with equations (\ref{eq:fastest_growing_wavelength_sech2}) and (\ref{eq:fastest_growing_wavelength_sech2_z0}) the relation
\begin{equation}
L_{\mathrm{sech^2}} = A_{\mathrm{sech^2}} \;  L_{0} = A_{\mathrm{sech^2}} \; \frac{ c_{\mathrm{s}}^{2}}{G \Sigma}.
\label{eq:clump_radius_S}
\end{equation}
From this we can conclude that in thick discs, the initial radial ring widths are much larger than the scale-height and than in the razor-thin limit assumed. \\
The total mass of a ring can be estimated to first order as
\begin{equation}
M_{\mathrm{Ring}} = \upi \Sigma(R) \; \left[ \left( R + \frac{ \lambda_{f_{\mathrm{sech^2}}}}{2}\right)^{2} - \left( R - \frac{\lambda_{f_{\mathrm{sech^2}}}}{2}\right)^{2} \right],
\end{equation}
which gives
\begin{equation}
M_{\mathrm{Ring}} = 2 \upi \Sigma (R) \; R \; \lambda_{f_{\mathrm{sech^2}}} ,
\end{equation}
where $R$ is the location of the maximum of the density perturbation.

\begin{figure*}
\centering
\subfloat[ \label{fig:surface_density}]
  {\includegraphics[width=.33\linewidth]{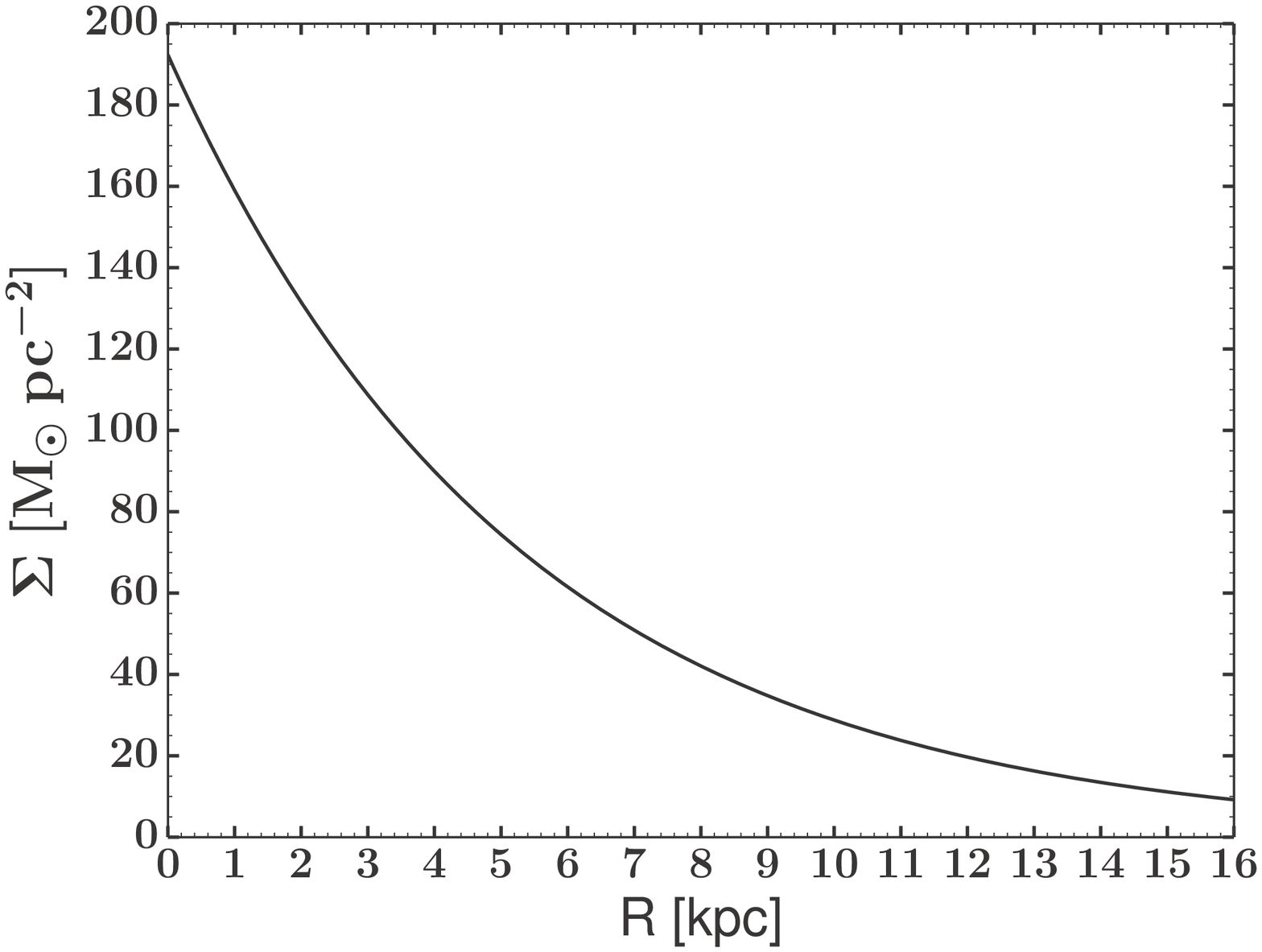}}\hfill
\subfloat[ \label{fig:scale_height}]
  {\includegraphics[width=.33\linewidth]{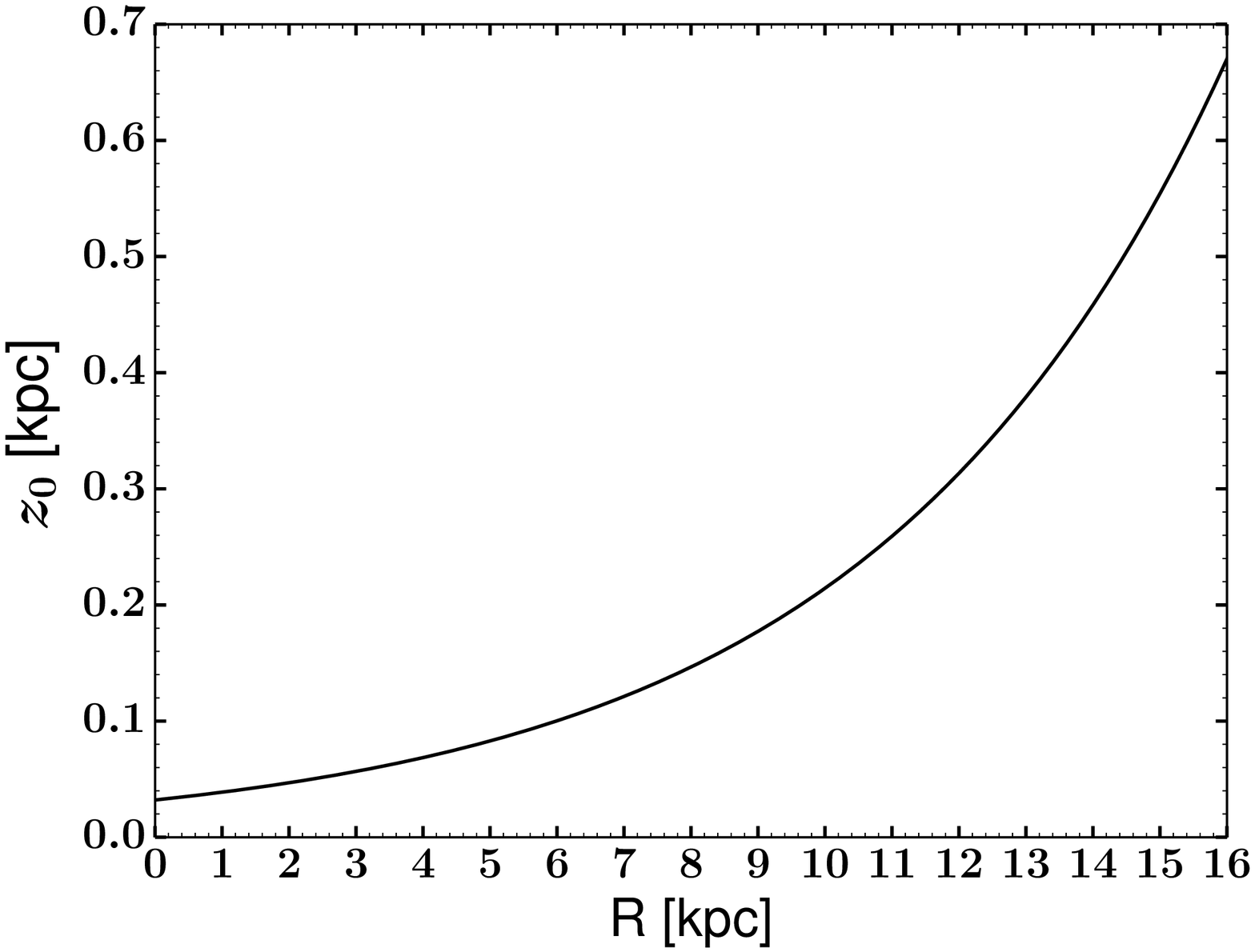}}\hfill
\subfloat[ \label{fig:rotation_curve}]
  {\includegraphics[width=.33\linewidth]{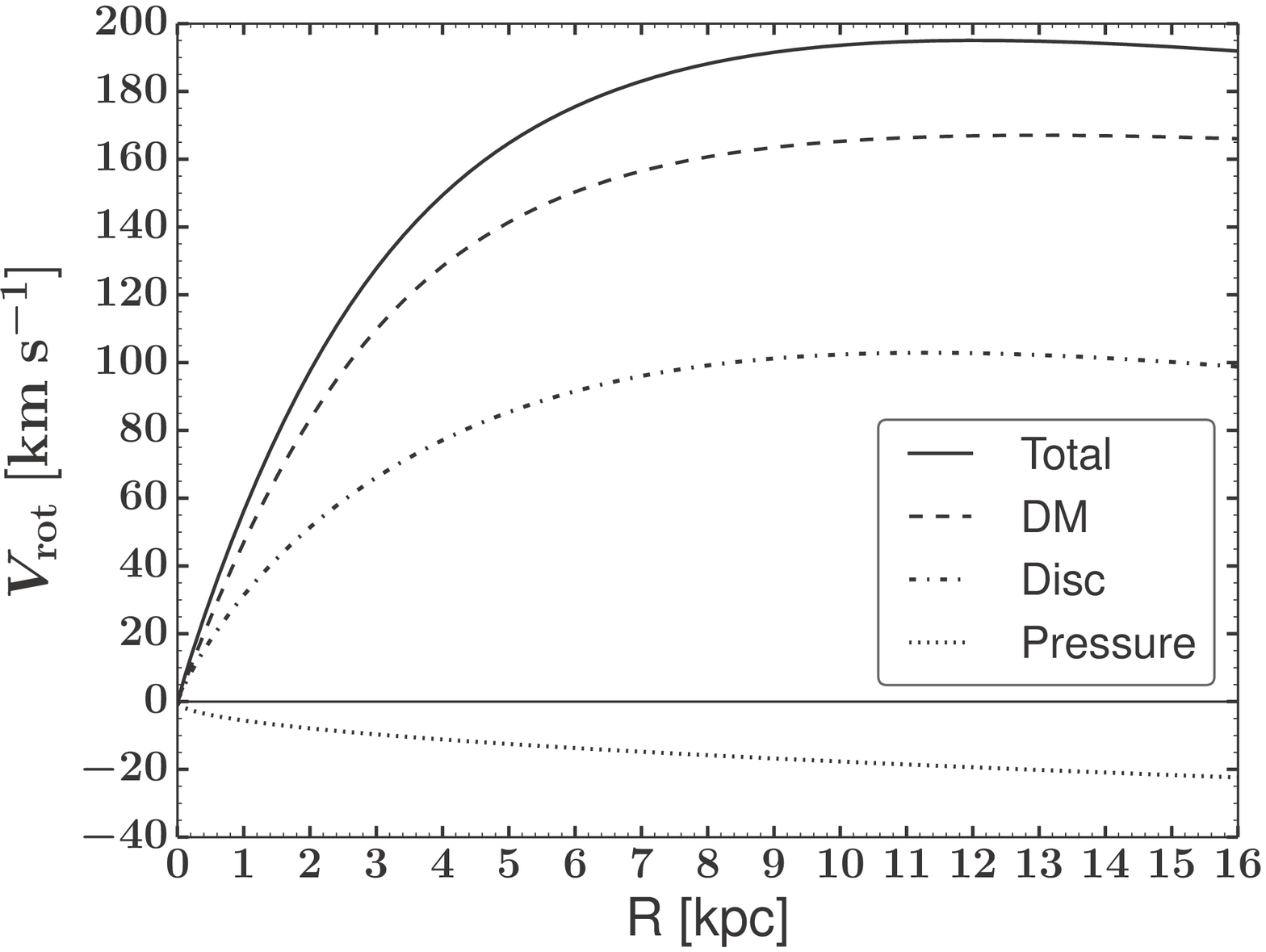}} 
\caption{The radial setup properties of the gas disc with the (a) declining exponential surface density profile $\Sigma(R)$ (equation \ref{App:first_requirenment}), (b) the increasing scale-height $z_0(R)$ (equation \ref{App:scale_height}), (c) the total rotation curve $V_{\mathrm{rot}}(R)$ (see Section \ref{subsec:Disk model}) and its decomposition into the contribution of the dominating dark matter halo, the gas disc and its correction due to the pressure gradient.}
\end{figure*}
\subsection{Local disc instability parameter}
\label{subsec:Local disc instability parameter}
Here we derive a local dimensionless stability parameter for a disc with finite thickness, following the example of \citet{1964ApJ...139.1217T} for an infinitesimal-thin gas disc with
\begin{equation}
\label{eq:Toomre_Q}
 Q_0 = \frac{\kappa \; c_{\mathrm{s}}}{\upi G \Sigma}.
\end{equation}
In contrast to other derivations we, apply the fastest growing wavelength on the dispersion relation to determine the parameter $Q$. \\
The disc is locally unstable for axisymmetric instabilities if $Q_0 < 1$. $Q_0$ can be obtained from the unmodified dispersion relation $\omega_0^2$, for equation (\ref{eq:modified_dispersion_relation_generalized}) with $z_0 = 0$. The global minimum of the dispersion relation determines the fastest growing $\lambda_{f_0}$ (equation \ref{eq:fastest_growing_wavelength_thin_disk}) and therefore is the 'last' possible wavelength which could grow if an unstable disc turns over to stability (from $\omega^2 < 0$ to $\omega^2 \geq 0$). The transition is given by $\omega_0^2(\lambda_{f_0}) = 0$ and by rearranging, it leads to the expression given in equation ($\ref{eq:Toomre_Q}$) with $Q_0 = 1$. In a similar way we proceed with the modified dispersion relation (equation \ref{eq:modified_dispersion_relation_generalized}) for the $\sech^2$ density profile and with the fastest growing wavelength we have
\begin{equation}
\label{eq:disp_rel_fastest_grow_wavelength_sech2}
\omega_{\mathrm{sech^2}}^2 = \kappa^2 - \frac{4 \upi^2 G \Sigma}{\lambda_{f_{\mathrm{sech^2}}}} F_{\mathrm{sech^2}} + \frac{4 \upi^2 c_{\mathrm{s}}^2}{\lambda_{f_{\mathrm{sech^2}}}^2} = 0.
\end{equation} 
We insert $\lambda_{f_{\mathrm{sech^2}}}$ (equation \ref{eq:fastest_growing_wavelength_sech2}) to get
\begin{equation}
\omega_{\mathrm{sech^2}}^2 = \kappa^2 -  \frac{\upi^2 G^2 \Sigma^2}{c_{\mathrm{s}}^2} \left( \frac{2 \; F_{\mathrm{sech^2}}}{A_{\mathrm{sech^2}}} - \frac{1}{A_{\mathrm{sech^2}}^2} \right)   = 0.   
\end{equation}
Now, we substitute the classical parameter $Q_0$ (equation \ref{eq:Toomre_Q}) and get
\begin{equation}
\label{eq:Q_sech2}
Q_{\mathrm{sech^2}} = Q_0 \times C_{\mathrm{sech^2}} = 1,
\end{equation}
with the constant proportionality factor
\begin{equation}
C_{\mathrm{sech^2}} = \frac{A_{\mathrm{sech^2}}}{\sqrt{2 \; F_{\mathrm{sech^2}} \; A_{\mathrm{sech^2}} - 1}} \simeq 1.437.
\end{equation} 
There is thus a simple linear relation between the thin disc approximation and a disc with finite thickness.
The error of using $Q_0 = 1$ is $30.411 \%$. The line of neutral stability, $Q_{\mathrm{sech^2}} = 1$, corresponds to $Q_{0, {\mathrm{crit}}} \simeq 0.696$, which is very similar to the approximation from \citet{Wang:2010wr} with $Q_{0, {\mathrm{crit}}} \simeq 0.693$. \\
For the exponential profile approximation we have $C_{\mathrm{exp}} = 1.546$ and hence an error of $7.585\%$ compared to $C_{\mathrm{sech^2}}$. In this case, the critical value is $Q_{0, {\mathrm{crit}}} \simeq 0.647$, consistent with what is found by \citet{Kim:2002gl}.
\subsection{Time-scales}
\label{subsec:Timescales}
To get an estimation of the fragmentation time-scales, we calculate the time at which the amplitude of the fastest growing disturbance has increased by a factor of $\bmath{e}$. For negative $\omega^2$, the growth rate $p$  is $- p^2 = \omega^2$ and the amplitude of a perturbation grows with $\exp(p \; t)$ \citep{Binney:2011vb}. For an increase by a factor of $\bmath{e}$, the growth time-scale is $t = 1 / p$. For the thin disc approximation,
\begin{equation}
t_{0} =\left(  \frac{ \upi^{2} G^{2} \Sigma^{2}}{ c_{\mathrm{s}}^{2}}  - \kappa^{2} \right) ^{-1/2},
\end{equation}
and expressed by the $Q_0$ parameter
\begin{equation}
\label{eq:t_0} 
t_{0} =\kappa^{-1} \left(  \frac{ 1}{ Q_{0}^{2}} - 1 \right) ^{-1/2}.
\end{equation}
By taking the thickness with the $\sech^2$ profile into account, we get the general form
\begin{equation}
t_{\mathrm{sech^2}} = \left(  \frac{4 \upi^2 G \Sigma}{\lambda_{f_{\mathrm{sech^2}}}} F_{\mathrm{sech^2}} - \frac{4 \upi^2 c_{\mathrm{s}}^{2}}{\lambda_{f_{\mathrm{sech^2}}}^{2}} - \kappa^{2} \right) ^{-1/2},
\end{equation}
and by substituting $\lambda_{f_{\mathrm{sech^2}}}$ with equation (\ref{eq:fastest_growing_wavelength_sech2}) and inserting $Q_{0}$ it leads to
\begin{equation}
\label{eq:t_sech^2}
t_{\mathrm{sech^2}} =\kappa^{-1} \left(  \frac{ 1}{ Q_{0}^{2} \; C_{\mathrm{sech^2}}^2}  - 1 \right) ^{-1/2},
\end{equation}
which goes to infinity for a marginally stable disc.

\section{Numerical modelling}
\label{sec:Numerical modelling}
In order to test the predictions of the linear stability analysis, we employ simulations of an idealized gas disc. In the following, we describe the code and the disc model.

\subsection{The simulation code}
\label{subsec:The simulation code}
We use the hydrodynamical AMR (adaptive mesh refinement) code \textsc{ramses} \citep{2002A&A...385..337T} to perform simulations of a self-gravitating isolated gas disc with an isothermal equation of state (EoS), embedded in a dark matter halo. Since we are interested in the early phases without any strong discontinuities, here the Euler equations are being solved with the local Lax--Friedrichs scheme. The dark matter is handled as a static external density field added to the source term in the Poisson solver. \\
The mesh in the 48-kpc simulation box is structured by nested AMR grids from the coarsest level of 187.5 pc outside the disc to a maximum resolution of $\Delta x_{\mathrm{min}}= 5.86$ pc inside, for the densest regions, employing an effective resolution of $2^8$--$2^{13}$ grid cells in all three direction. At each resolution level, we ensure that the Jeans length is resolved by at least 18 grid cells (see Section \ref{subsec:Numerical considerations}).
\begin{figure}
\includegraphics[width=84mm]{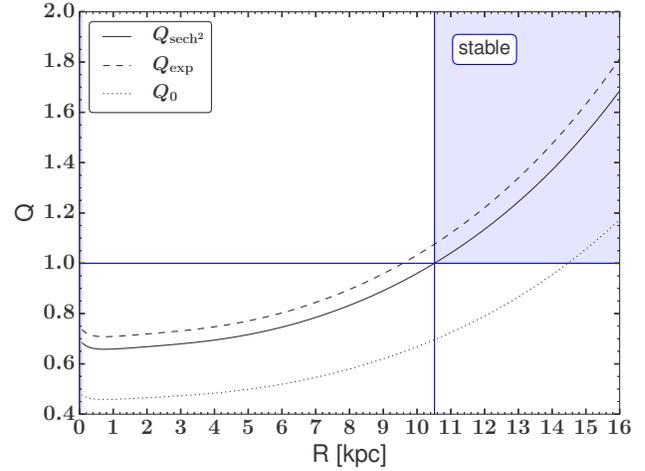}
\caption{Local disc instability parameter: the solid line represents the numerical solution, according to the vertical $\sech^2$ density profile (equation \ref{eq:Q_sech2}). The disc is unstable between radius $R = 0-10.517$ kpc and the stable part is illustrated by the blue shaded region.  The exponential profile approximation (dashed line), gives a slightly more stable disc and therefore results in a smaller unstable region (see Section \ref{subsec:Local disc instability parameter}). The thin disc approximation, for $z_0 = 0$, with Toomre's $Q_0$ (dotted line) (equation \ref{eq:Toomre_Q}), overestimates the instability dramatically, in intensity and with respect to the extent of the unstable region.
\label{Q}} 
\end{figure}

\begin{figure*}
\centering
\subfloat[\label{Results:fig:theory:growing_wavlength}]
  {\includegraphics[width=.49\linewidth]{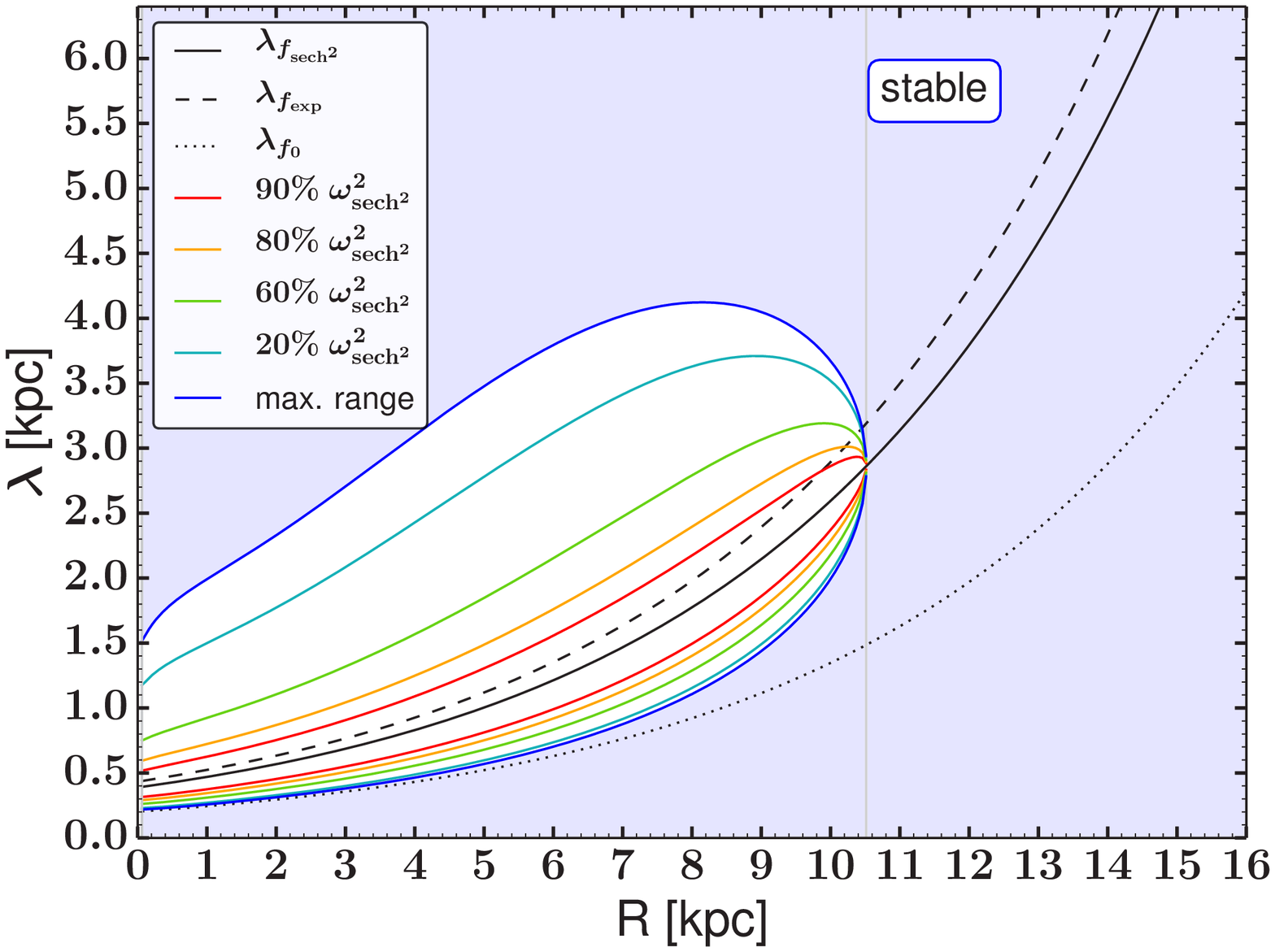}}\hfill
\subfloat[ \label{Results:fig:theory:growth_rates_wavelength}]
  {\includegraphics[width=.49\linewidth]{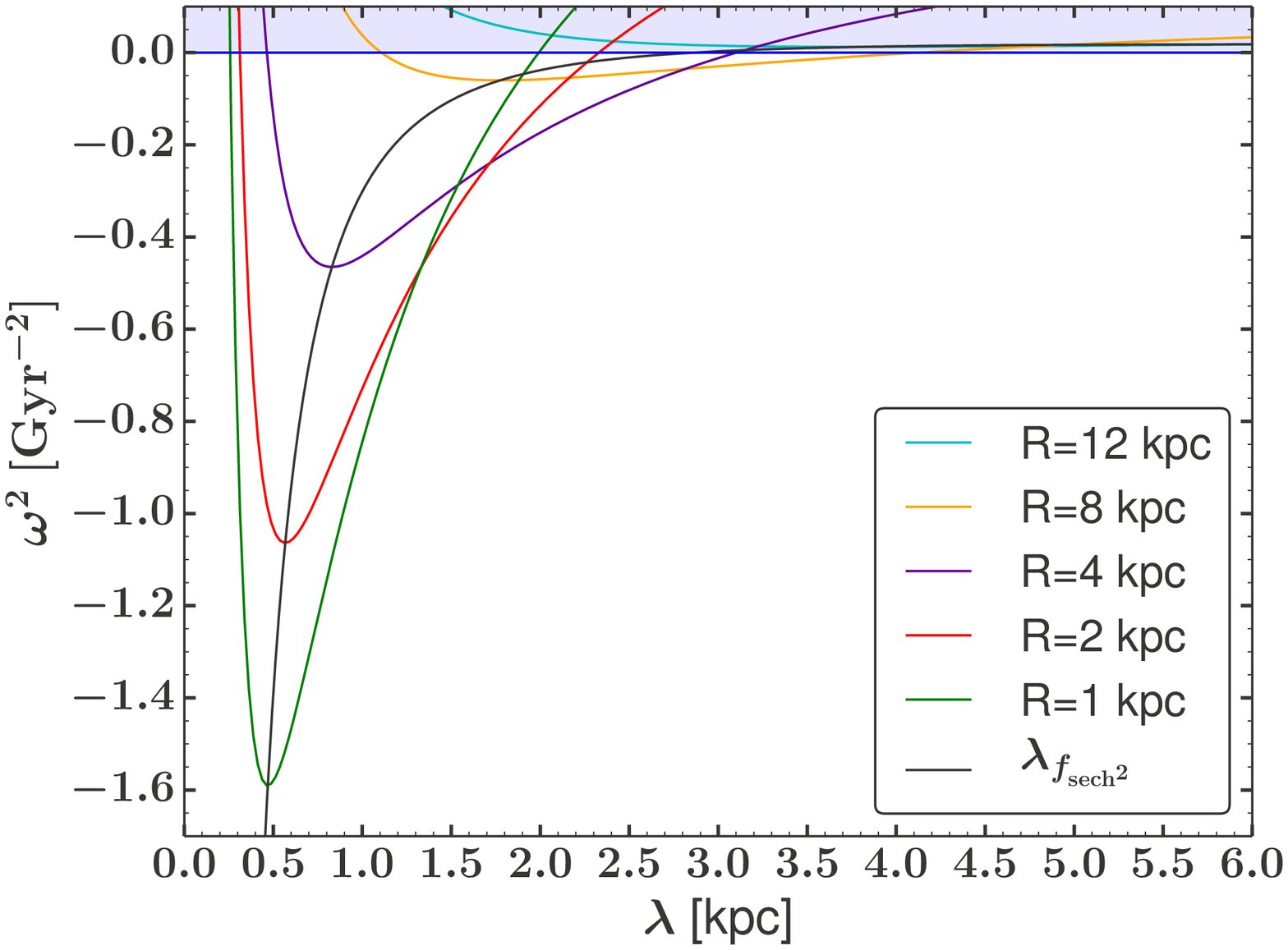}}\hfill
\caption{(a) Possible growing wavelength at every disc radius. The black lines correspond to the fastest growing perturbation wavelength, of the $\sech^2$ density profile (solid) (equation \ref{eq:fastest_growing_wavelength_sech2}), the exponential density approximation (dashed) (equation \ref{eq:fastest_growing_wavelength_exp}) and the razor-thin disc treatment (dotted) (equation \ref{eq:fastest_growing_wavelength_thin_disk}). The coloured solid lines represent unstable wavelengths with growth rates $p_{\mathrm{sech^2}} = \sqrt{- \omega^2_{\mathrm{sech^2}}}$ (Section \ref{subsec:Timescales}) that are a fraction of the fastest growth rate and the white area is the complete region of possible growing perturbations. The transition to the stable regime  (blue shaded region), with zero growth, is given by the blue solid lines. (b) The dispersion relation at different radii (coloured lines) gives the growth rate as a function of wavelength and represents slices of (a). The minimum of each $R$ gives the fastest growing perturbation wavelength with $\lambda_{f_{\mathrm{sech^2}}}$ at that radius (black line).}
\label{Results:fig:growing_wavlength__and__growth_rates}
\end{figure*}

\subsection{Disc model}
\label{subsec:Disk model}
We perform simulations of a massive gas disc $M_{\mathrm{disc}}=2.7 \times 10^{10} \; \mathrm{M_{\sun}}$  embedded in a spherical dark matter halo that within $16$ kpc has a mass of $M_{\mathrm{DM}} = 1.03 \times 10^{11} \; \mathrm{M_{\sun}}$ following the \citet{Burkert:1995jr} density profile with a scalelength $a_{0} = 4$ kpc. The gas temperature is $10^4$ K and the disc has an exponential surface density profile with scalelength $h = 5.26$ kpc and truncation at $R_{\mathrm{d}} = 16$ kpc (Fig. \ref{fig:surface_density}). The central density is $\rho_{\mathrm{c}} = 3 \; \mathrm{M_{\sun} \; pc^{-3}}$. The parameters were chosen to resemble an initially unstable massive high-redshift disc galaxy \citep{2011ApJ...733..101G}, with a relatively large scalelength, and a stable outer part with a relatively flat rotation curve.\\
The disc setup is initially in vertical hydrostatic equilibrium (see Appendix \ref{App:Hydrostatic equilibrium}), which naturally leads, for an isothermal disc, to an increasing scaleheight with radius (Fig. \ref{fig:scale_height}), see also \citet{Wang:2010wr}. Hydrodynamical equilibrium is achieved by following the steps described in \citet{Wang:2010wr}, with the total rotation curve as shown in Fig. \ref{fig:rotation_curve}, which consists of the contributions of the gas disc and the dark matter halo. For our $\sech^2$-profile, the disc is unstable for axisymmetric perturbations between $R_{\mathrm{u}}=0-10.517$ kpc (solid line in Fig. \ref{Q}), becoming more stable for larger radii and totally stable for $R \geq 10.517$ kpc. 
The exponential profile approximation assumes a slightly more stable disc with a smaller unstable regime $R_{\mathrm{exp}} = 0-9.58$ kpc (dashed curve in Fig. \ref{Q}). Using the classical Toomre $Q_0$, that is ignoring $z_0$, leads to a much more unstable disc between $R_0 = 0-14.47$ kpc (dotted curve in Fig. \ref{Q}).\\
To form gravitationally bound clumps, an efficient cooling is needed \citep{Gammie:2001hv,Dekel:2009bn}. This is ensured by the isothermal EoS, which keeps the temperature at $10^4$ K at all densities.

\subsection{Numerical considerations}
\label{subsec:Numerical considerations} 
To avoid artificial fragmentation in gravitationally collapsing gas, the Jeans length has to be resolved by at least four cells $N_{\mathrm{J}} = 4$ \citep{1997ApJ...489L.179T}.  \citet{Ceverino:2010eh} found convergence in their simulations in clump numbers and masses by resolving the Jeans length with at least seven elements, $N_{\mathrm{J}} = 7$, at each refinement level. \\
Furthermore, it is also crucial to resolve the mid-plane sufficiently, where the structures in the disc form first. Too low resolution cannot represent the higher densities there, which can lead to an unreasonable structure formation. Due to the Jeans length refinement and the density distribution of the disc, the  initial AMR grid has a resolution gradient, with smaller cells in the galactic centre and larger ones at larger radius and height. The scaleheight is represented with at most two cells for $N_{\mathrm{J}} = 7$,  and by increasing the number in test simulations, we found, that the ring-like structures emerge properly from the disc for $N_{\mathrm{J}} \geq 18$ grid cells per Jeans length, which corresponds to five cells per scaleheight at all radii. Too low resolution effectively raises the $Q$ of the disc numerically and spiral-like features appear, as expected for values $Q \geq Q_{\mathrm{crit}}$, where axisymmetric modes are stable but nonaxisymmetric modes can still grow. The isothermal EoS keeps the disc scaleheight constant and therefore ensures, that it is sufficiently resolved until the structures begin to grow.\\
The disc is isothermal with $10^4$ K. However to ensure the Jeans condition for higher densities also at maximum resolution, we add an artificial pressure floor \citep{2010MNRAS.409.1088B, Agertz:2009wd} 
\begin{equation} 
T \geq \frac{G \; m_{\mathrm{H}}}{\upi \; k_{\mathrm{B}} \gamma}  N^{2} \Delta x_{\mathrm{min}}^{2} \; \rho,
\end{equation}
where $m_{\mathrm{H}}$ is the atomic mass of hydrogen, $k_{\mathrm{B}}$ the Boltzmann constant, $\gamma = 1$ the adiabatic index, $\rho$ the density and $N=18$ the number of resolution elements per Jeans length for the smallest scales $\Delta x_{\mathrm{min}}$. 
This does not affect the global ring formation but determines the thickness and pressure in the collapsed high-density ring structures and by this regulates their fragmentation into clumps. The effect of such a pressure floor on the clump numbers, final sizes and their interactions are beyond the scope of this study and will be investigated in a subsequent paper.

\section{Results}
\label{sec:Results} 

\subsection{Perturbation theory}
\label{Results:subsec:Perturbation theory}
\begin{figure}
\includegraphics[width=84mm]{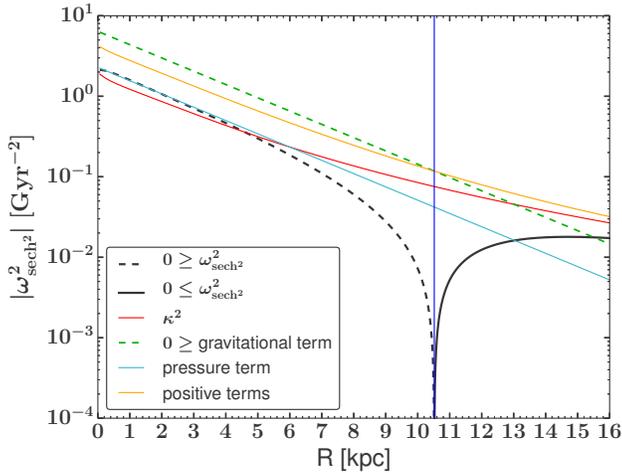}
\caption{Minimum of the dispersion relation $\omega^2_{\mathrm{sech^2}} = \omega^2 (\lambda_{f_{\mathrm{sech^2}}})$ (equation \ref{eq:modified_dispersion_relation_generalized}) in absolute values (black lines) and its decomposition (coloured lines) at every radius. The negative gravitational term (green) of equation (\ref{eq:modified_dispersion_relation_generalized}) dominates the unstable regime (black dashed line), while the positive contributions become more important in the stable regime (black solid line). The stabilising terms (orange) consist of the epicyclic frequency term (red), with $\kappa^2$ and the pressure term (cyan), with $c_{\mathrm{s}}^2 \left( \frac{2 \upi}{\lambda_{f_{\mathrm{sech^2}}}} \right)^2$. The effects of the epicyclic frequency and the pressure are comparable in the inner part of the disc, while $\kappa^2$ dominates for larger radii.
\label{Results:fig:dispersion_relation_1}}
\end{figure}

From the modified dispersion relation (equation \ref{eq:modified_dispersion_relation_generalized}) for the $\sech^2$ profile, we can derive the possible growing wavelengths for our disc model (Fig. \ref{Results:fig:theory:growing_wavlength}) in the unstable regime $R_{\mathrm{u}} = 0- 10.517$ kpc. The fastest growing perturbation wavelength is different for every radius and increases outwards. For the $\sech^2$ profile the range lies between $\lambda_{f_{\mathrm{sech^2}}} = [0.393, 2.862]$ kpc, while the exponential profile approximation would indicate $\lambda_{f_{\mathrm{exp}}} = [0.438, 3.191]$ kpc and in the razor-thin disc approximation we have $\lambda_{f_{0}} = [0.204, 1.486]$ kpc. For axisymmetric disturbances, the dispersion relation holds so long as $| k R | \gg 1$ \citep{Binney:2011vb}, which is $| 2 \upi R | \gg \lambda_{\mathrm{f_{sech^2}}}$ and here fulfilled for $R \gg 63$ pc and $\lambda_{\mathrm{f_{sech^2}}} \gg 392.7$. The growth rate is decreasing the larger the difference between $\lambda$ and $\lambda_{f_{\mathrm{sech^2}}}$ (Fig. \ref{Results:fig:growing_wavlength__and__growth_rates}) and reaches zero at the maximum range. At every radius there are two wavelengths with the same growth rate $p_{\mathrm{sech^2}} = \sqrt{- \omega^2_{\mathrm{sech^2}}}$, a smaller and a larger one, relative to $\lambda_{f_{\mathrm{sech^2}}}$. With decreasing growth rate the asymmetry in the difference between $\lambda$ and the corresponding fastest growing wavelength is increasing. In general, the smaller perturbation wavelengths in the disc centre can grow faster than the larger ones in the outer regime.\\
The dispersion relation $\omega^2$ (Section \ref{subsec:Exact fastest growing perturbation wavelength}) can be divided into a destabilizing term (negative), which is dominating the unstable regime $R_{\mathrm{u}}$, and stabilizing terms (positive), taking over in the stable region, see Fig. \ref{Results:fig:dispersion_relation_1}. The stabilizing contribution in the disc centre is due to pressure (expressed by $c_{\mathrm{s}}$) and differential rotation (expressed by $\kappa^2$), while going outwards, the epicyclic frequency is more important than pressure.
\subsection{General evolution of the surface density}
\label{Results:subsec:General evolution}
\begin{figure*}
\centering
\subfloat[ \label{fig:sim1}]
  {\includegraphics[width=.248\linewidth]{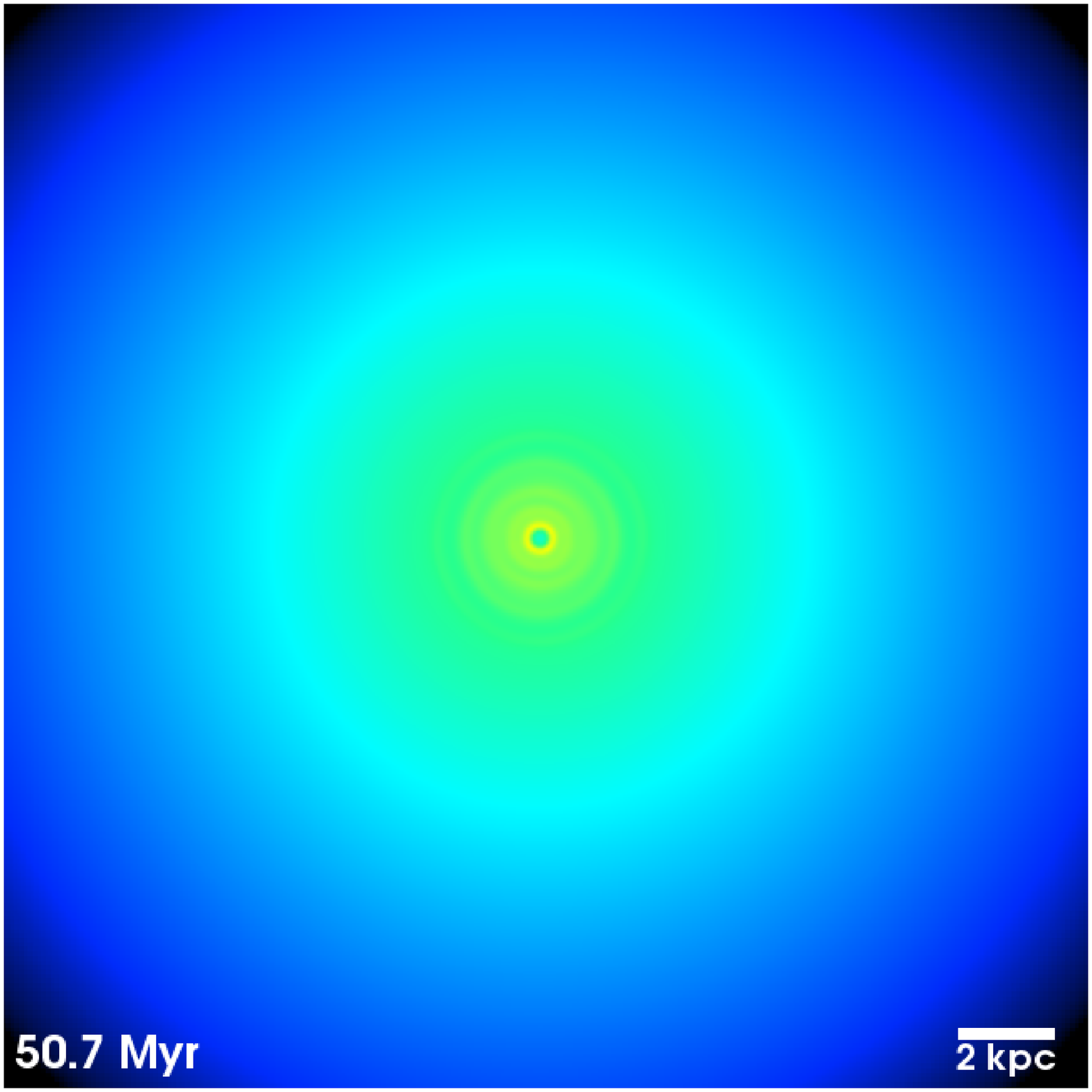}}\hfill
\subfloat[ \label{fig:sim2}]
  {\includegraphics[width=.248\linewidth]{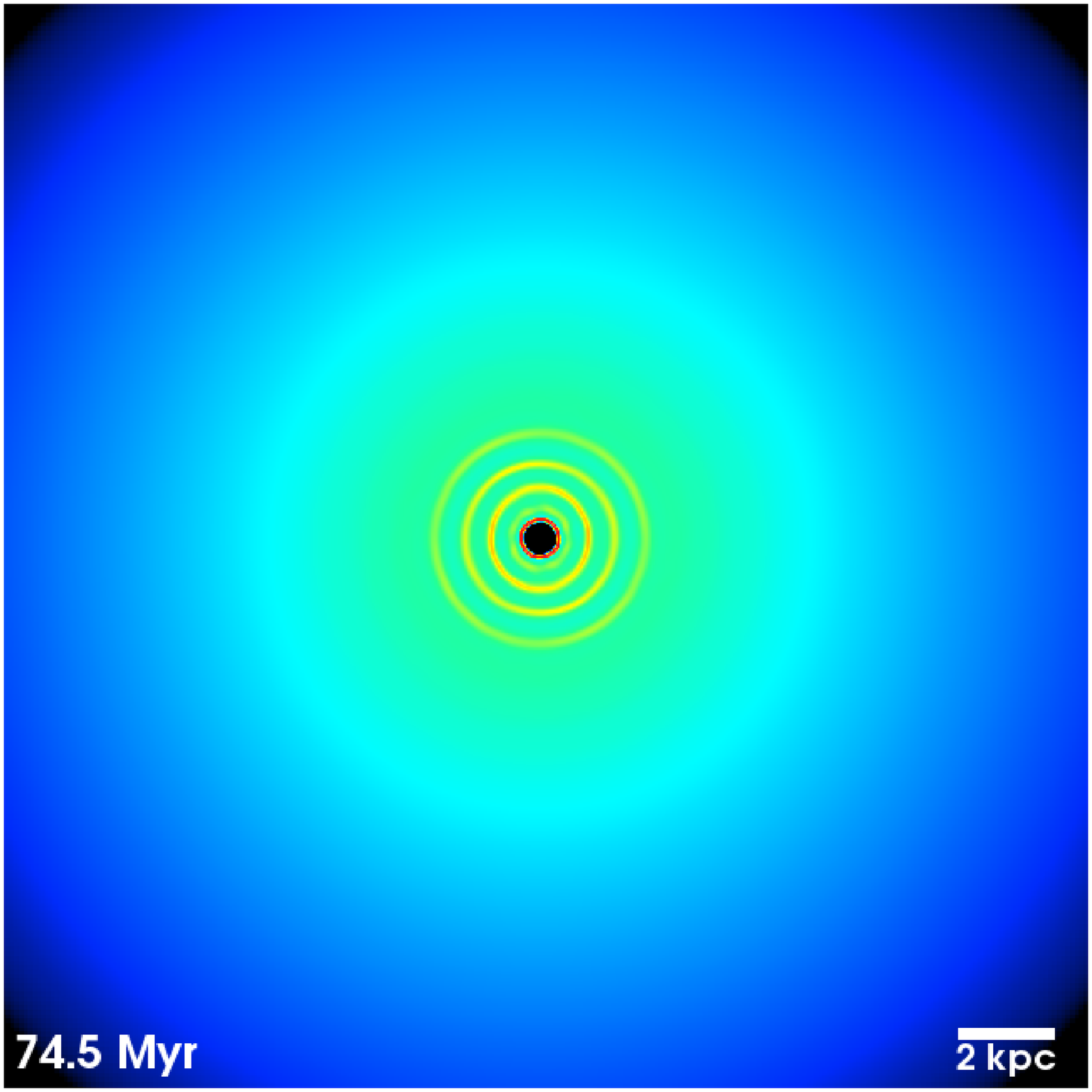}}\hfill
\subfloat[ \label{fig:sim3}]
  {\includegraphics[width=.248\linewidth]{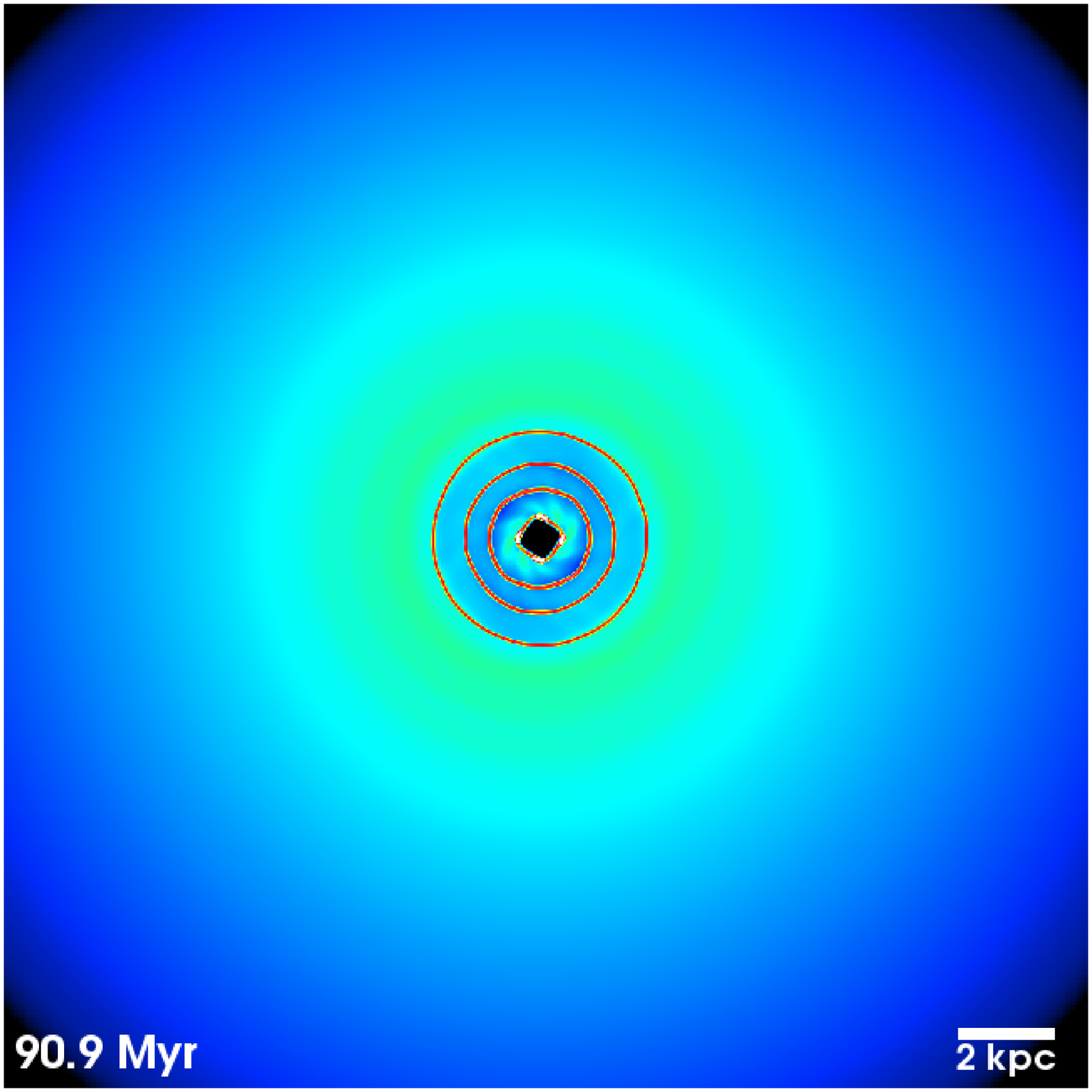}}\hfill
\subfloat[ \label{fig:sim4}]
  {\includegraphics[width=.248\linewidth]{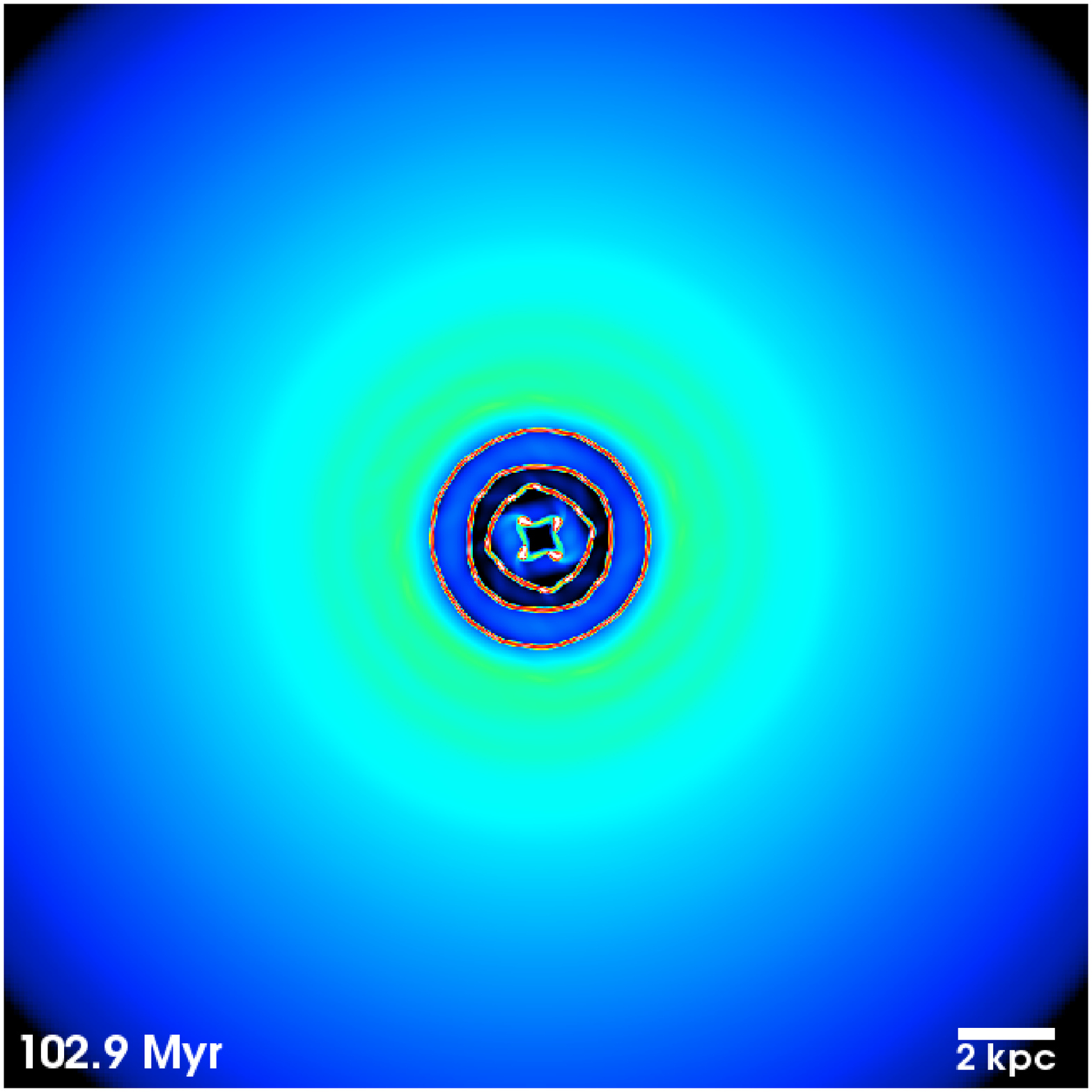}}

\subfloat[ \label{fig:sim5}]
  {\includegraphics[width=.248\linewidth]{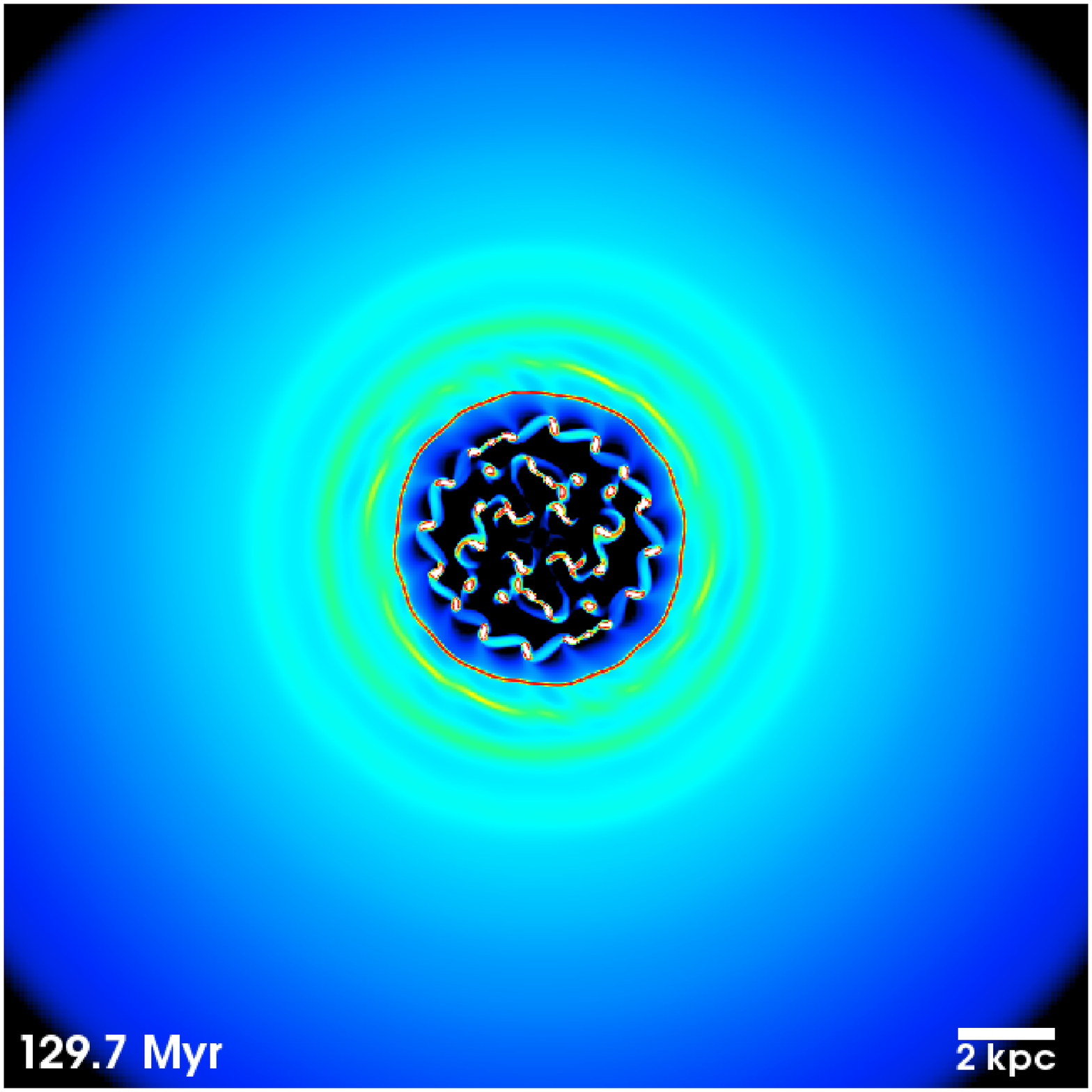}}\hfill
\subfloat[ \label{fig:sim6}]
  {\includegraphics[width=.248\linewidth]{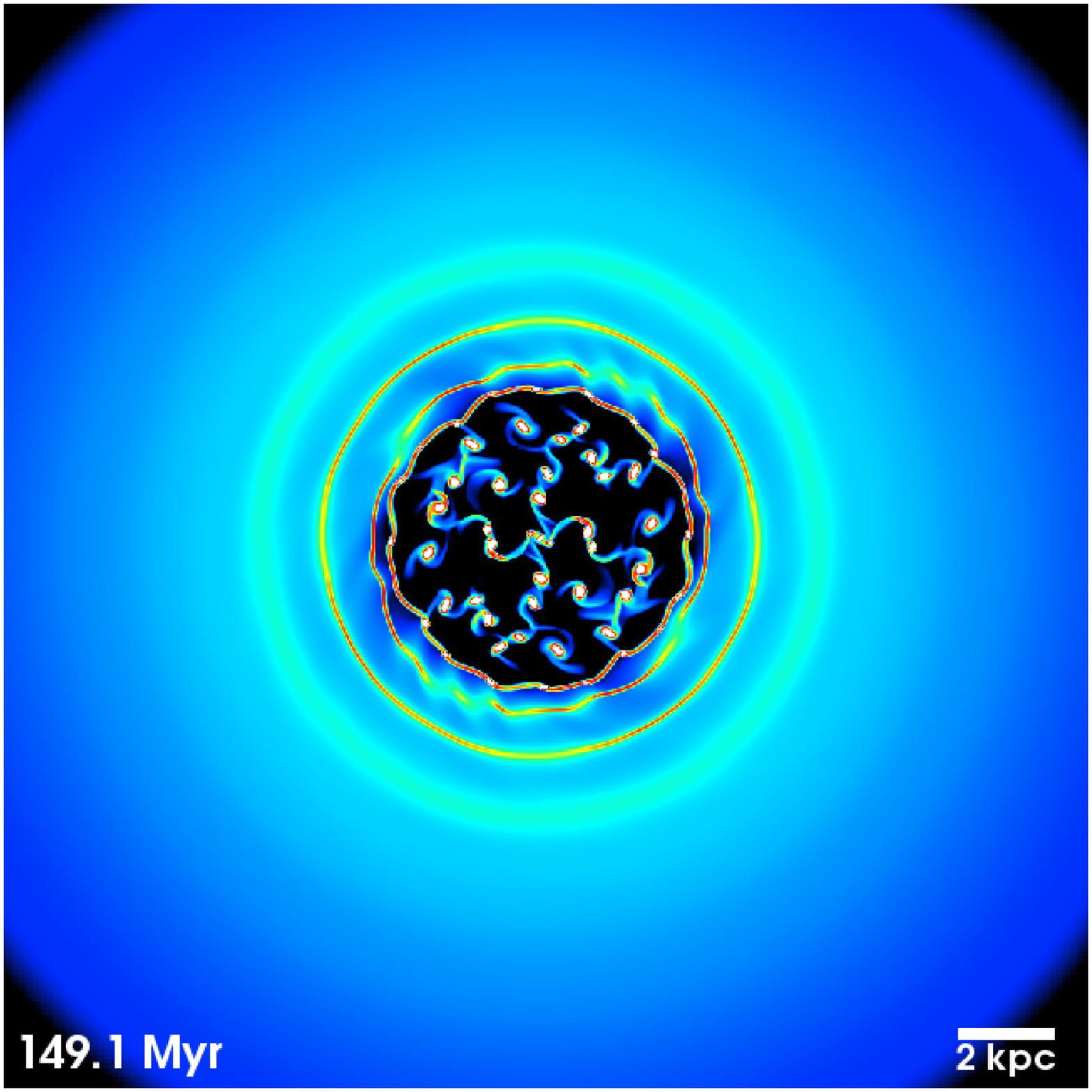}}\hfill
\subfloat[ \label{fig:sim7}]
  {\includegraphics[width=.248\linewidth]{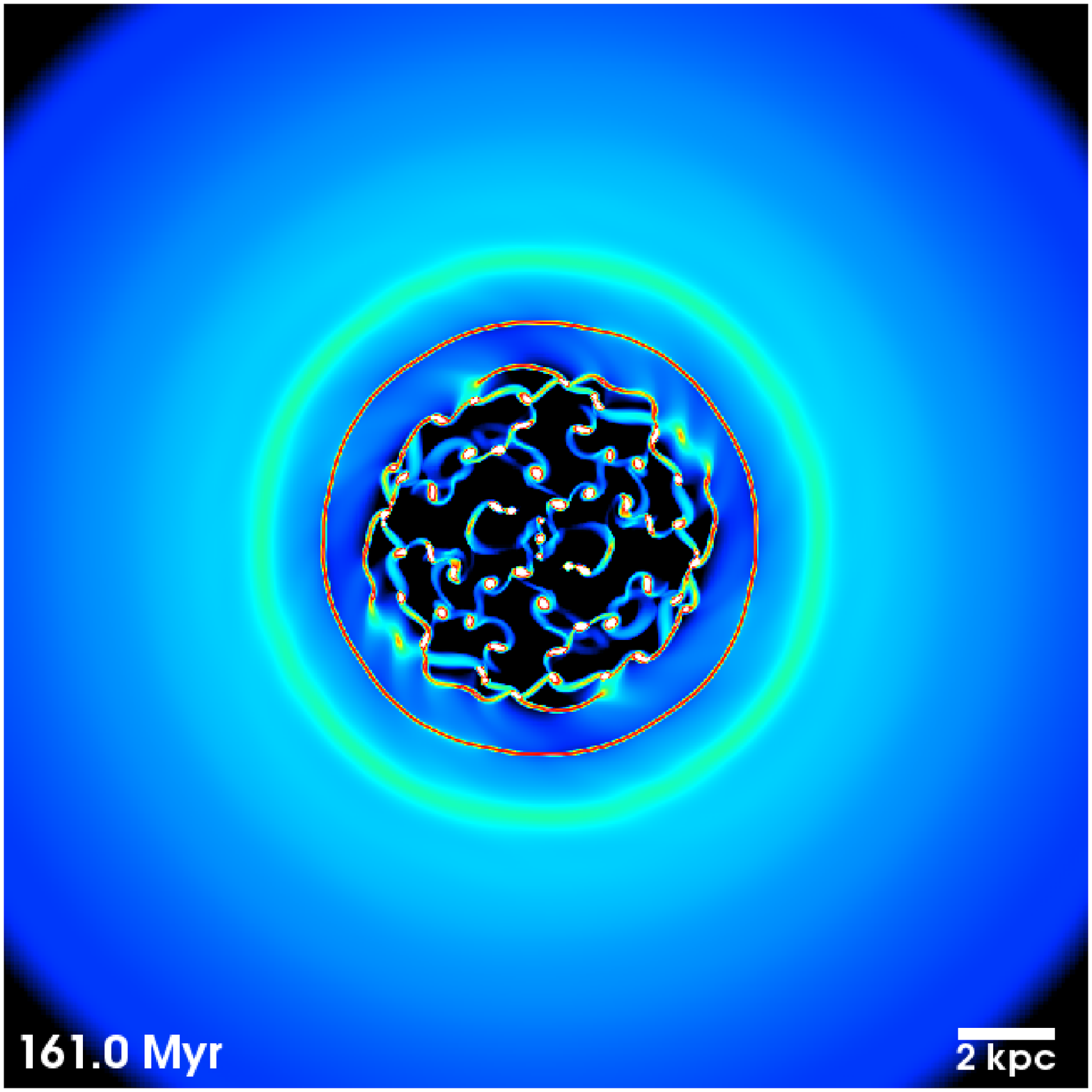}}\hfill
\subfloat[ \label{fig:sim8}]
  {\includegraphics[width=.248\linewidth]{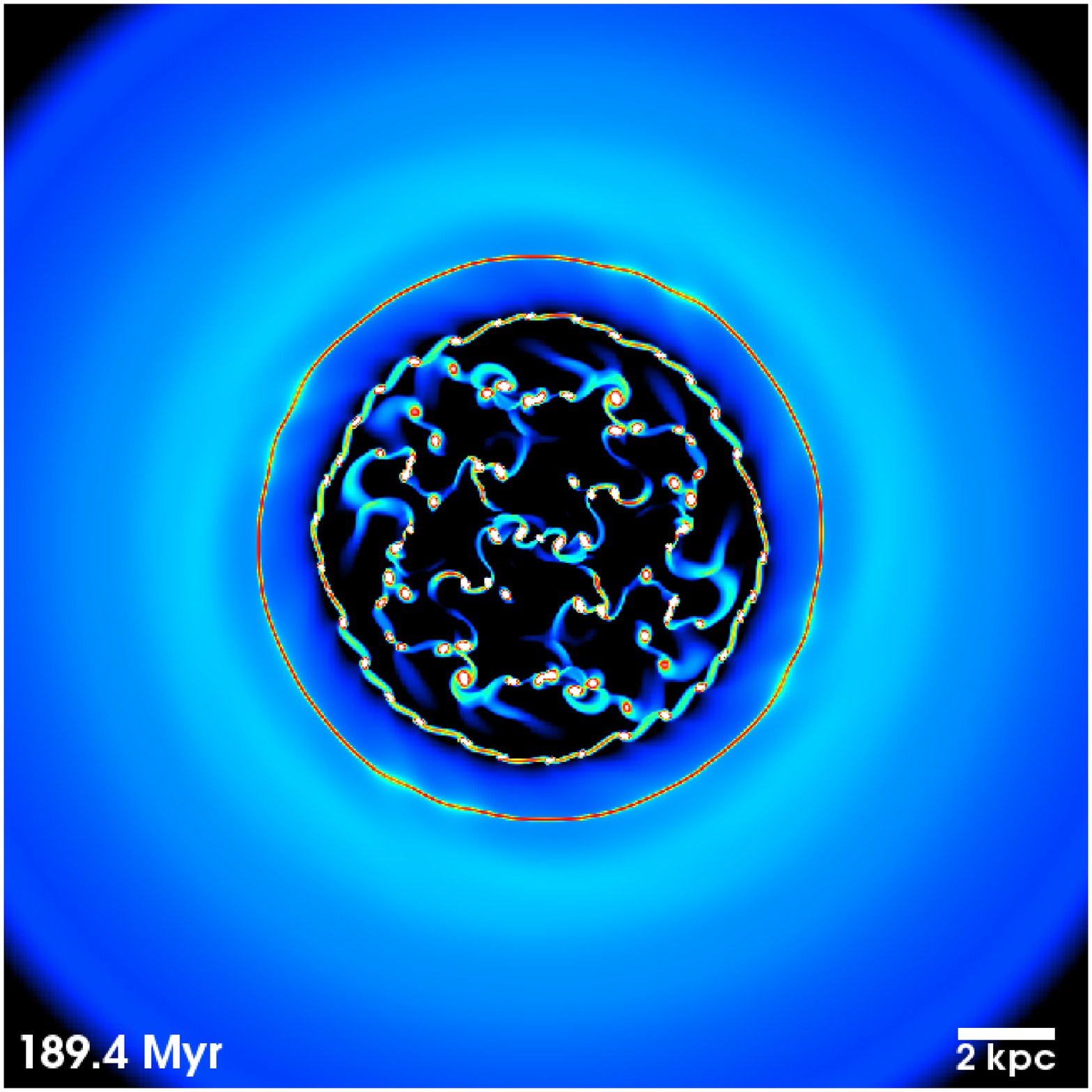}}

\subfloat[ \label{fig:sim9}]  
  {\includegraphics[width=.248\linewidth]{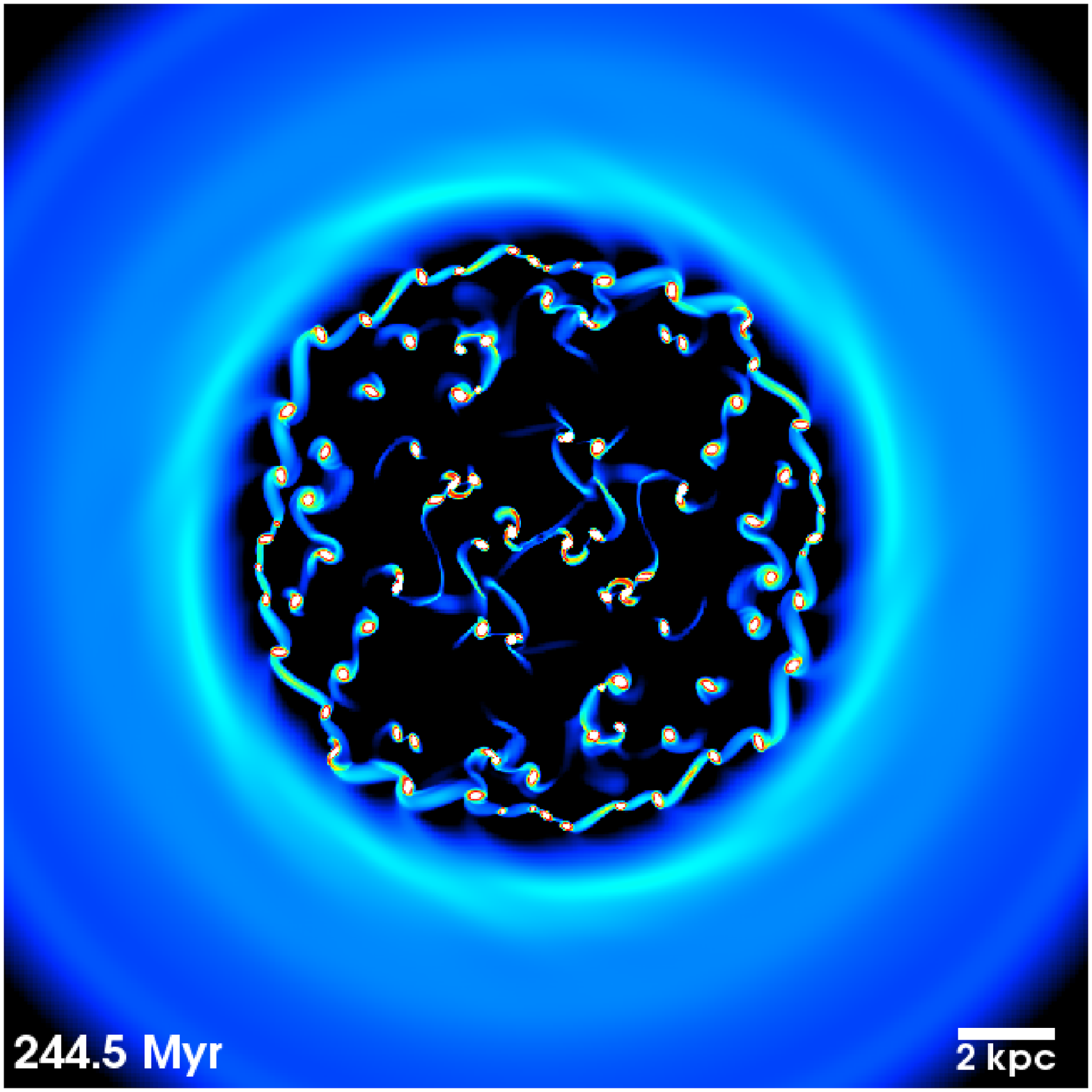}}\hfill
\subfloat[ \label{fig:sim10}]
  {\includegraphics[width=.248\linewidth]{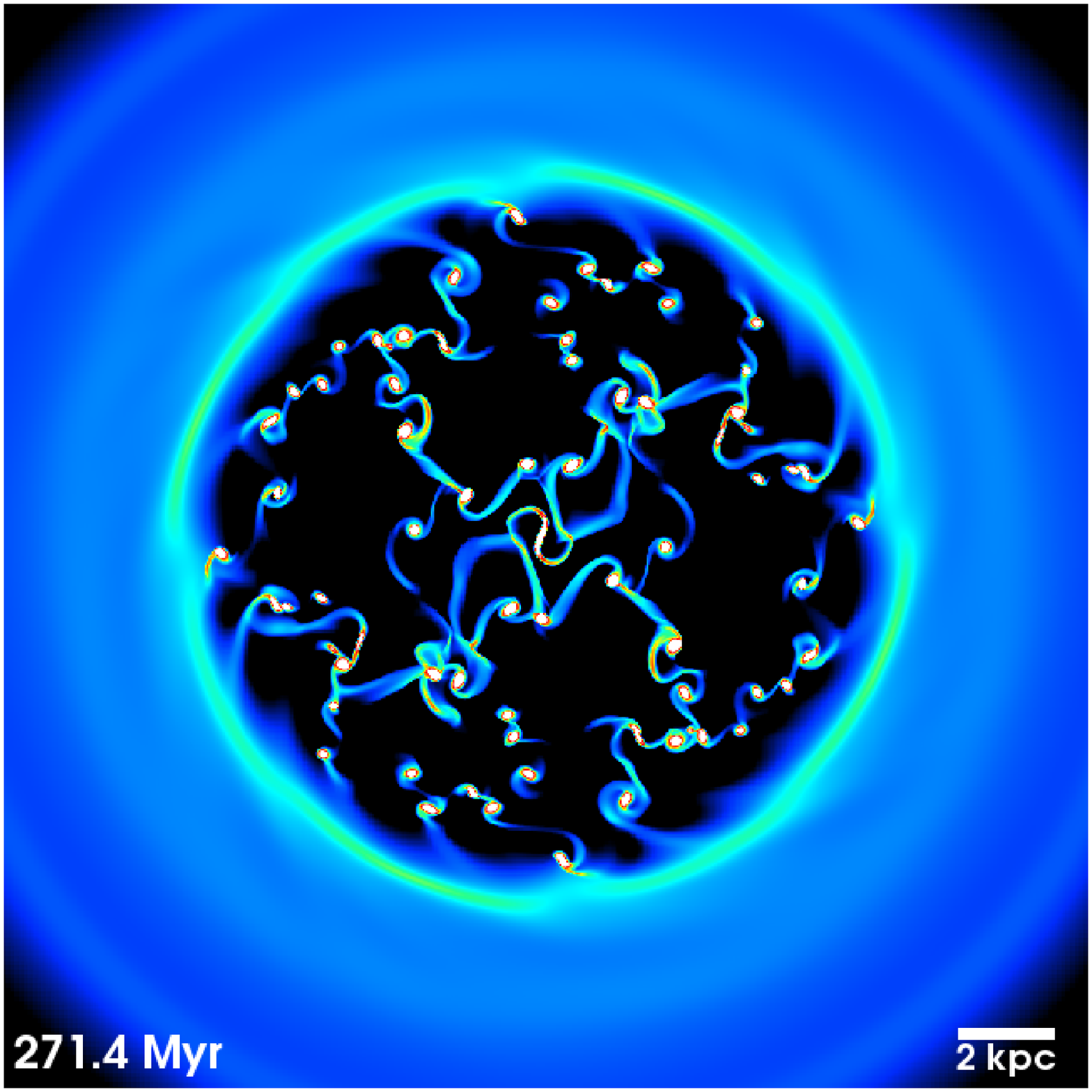}}\hfill
\subfloat[ \label{fig:sim11}]
  {\includegraphics[width=.248\linewidth]{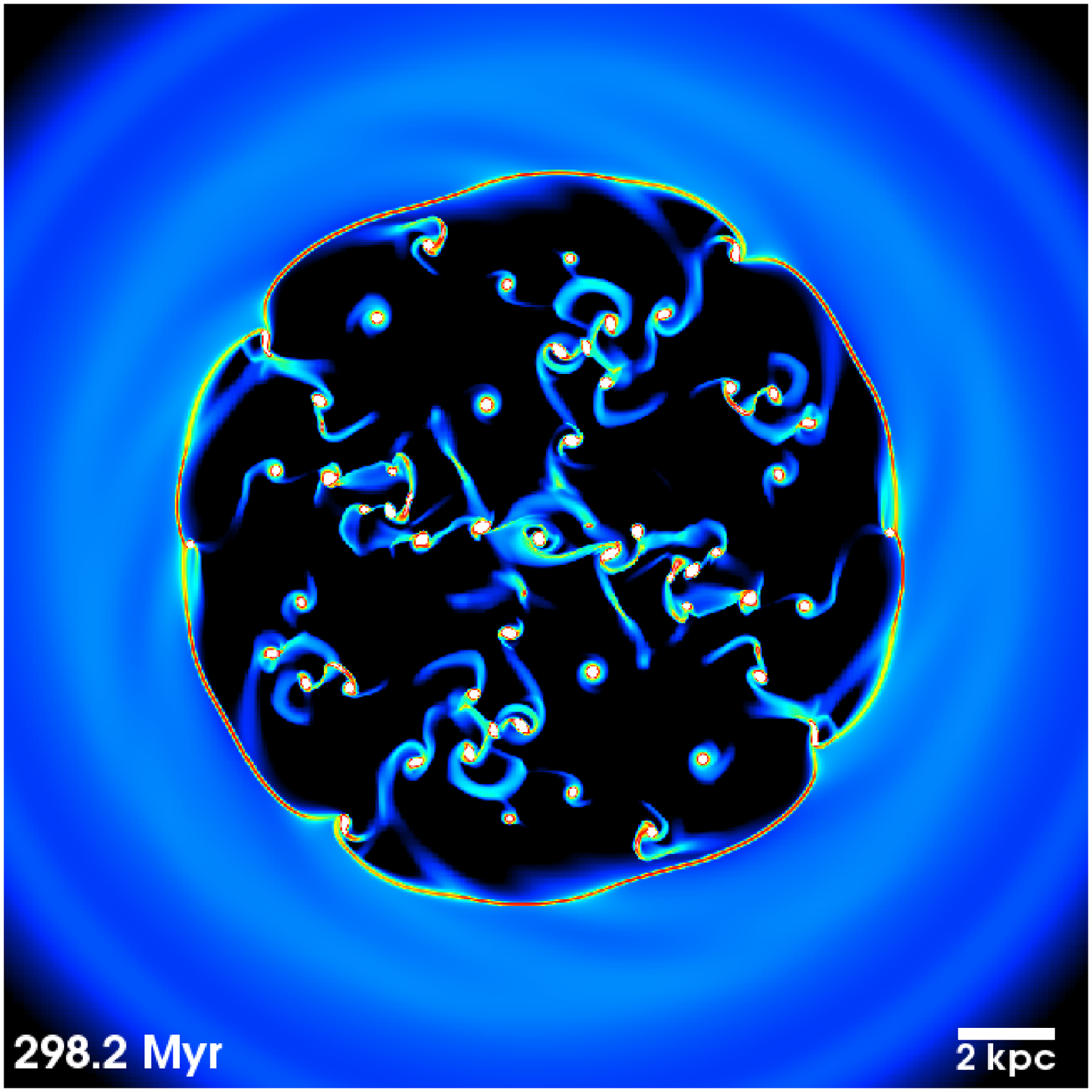}}\hfill
\subfloat[ \label{fig:sim12}]
  {\includegraphics[width=.248\linewidth]{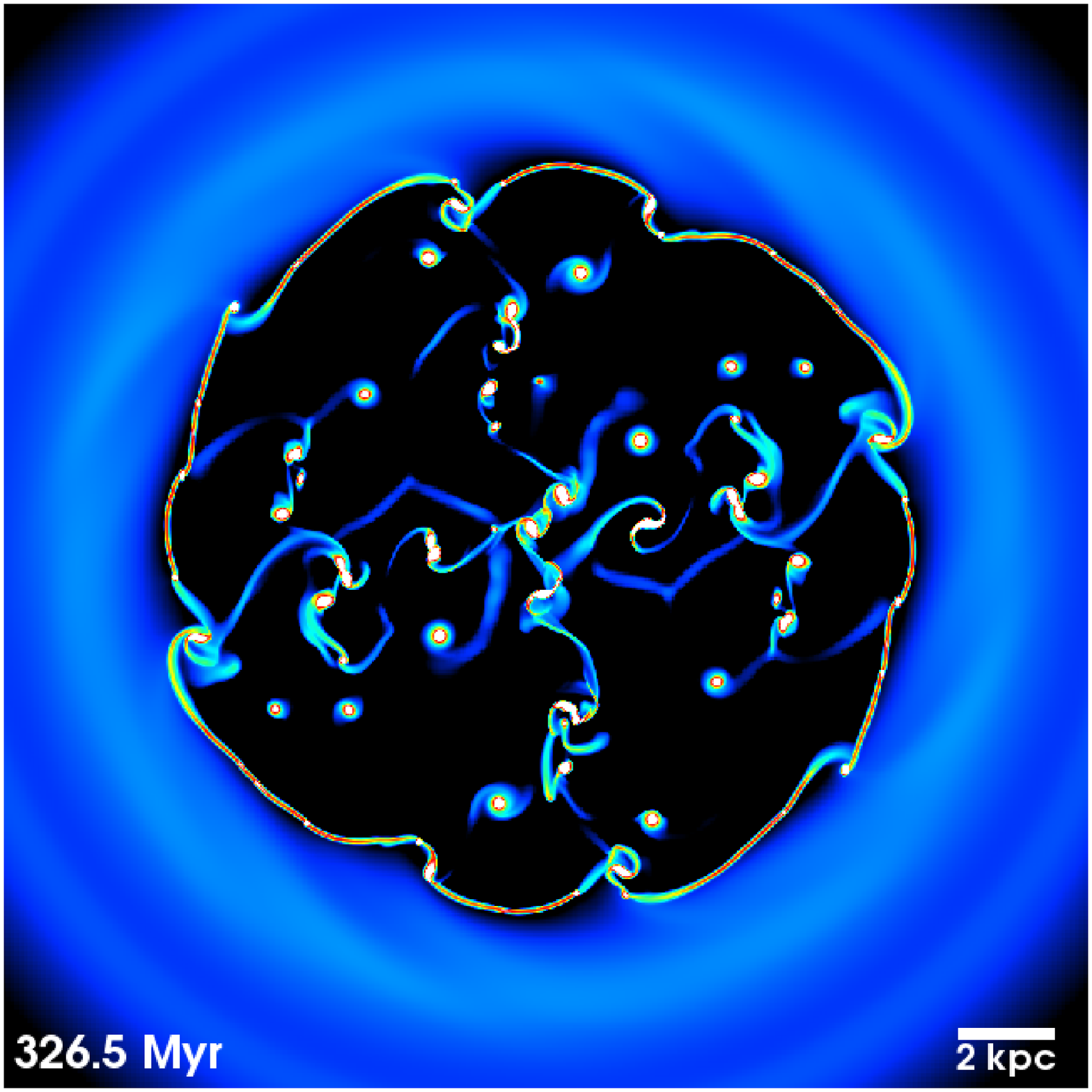}}  

\subfloat[ \label{fig:sim13}]  
  {\includegraphics[width=.248\linewidth]{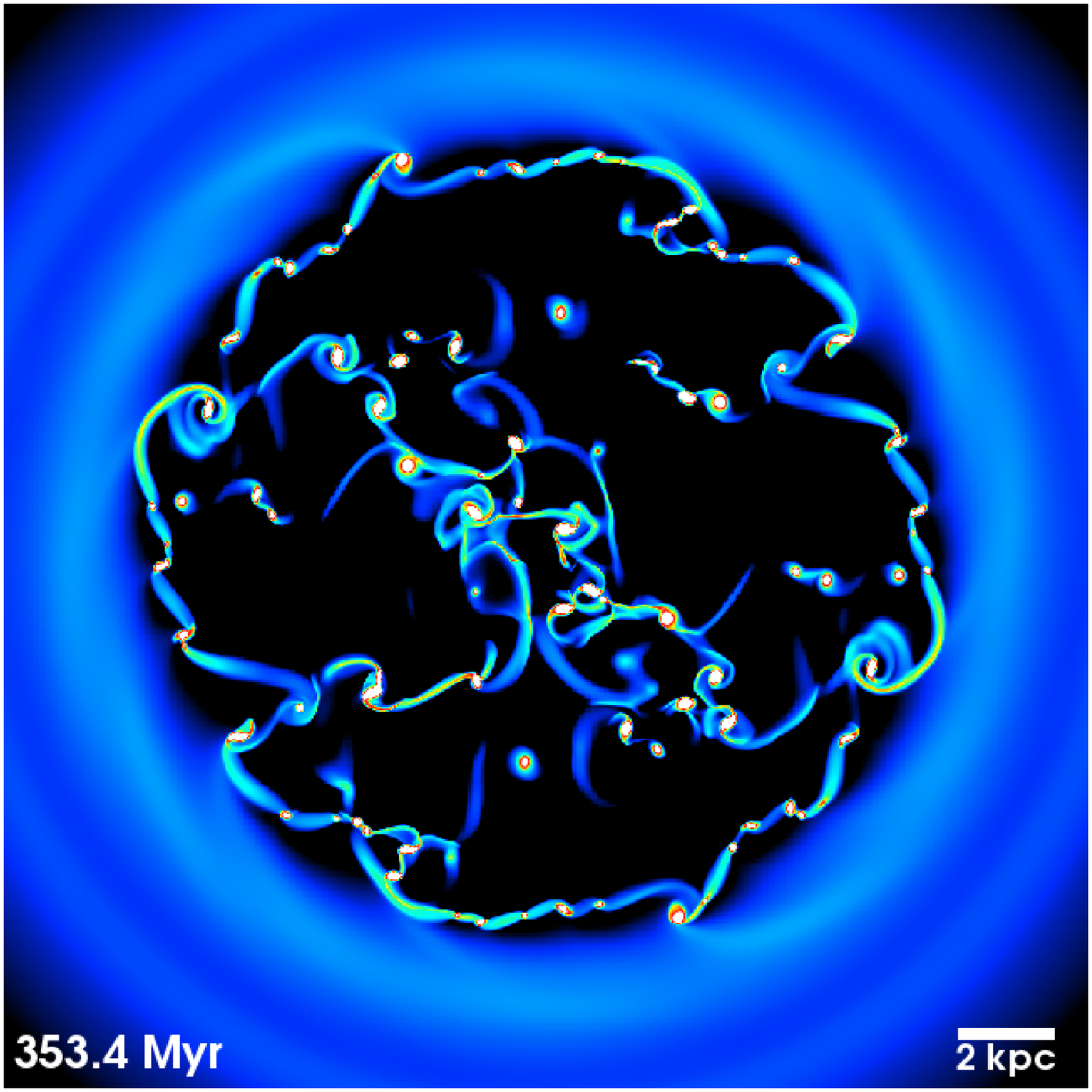}}\hfill
\subfloat[ \label{fig:sim14}]
  {\includegraphics[width=.248\linewidth]{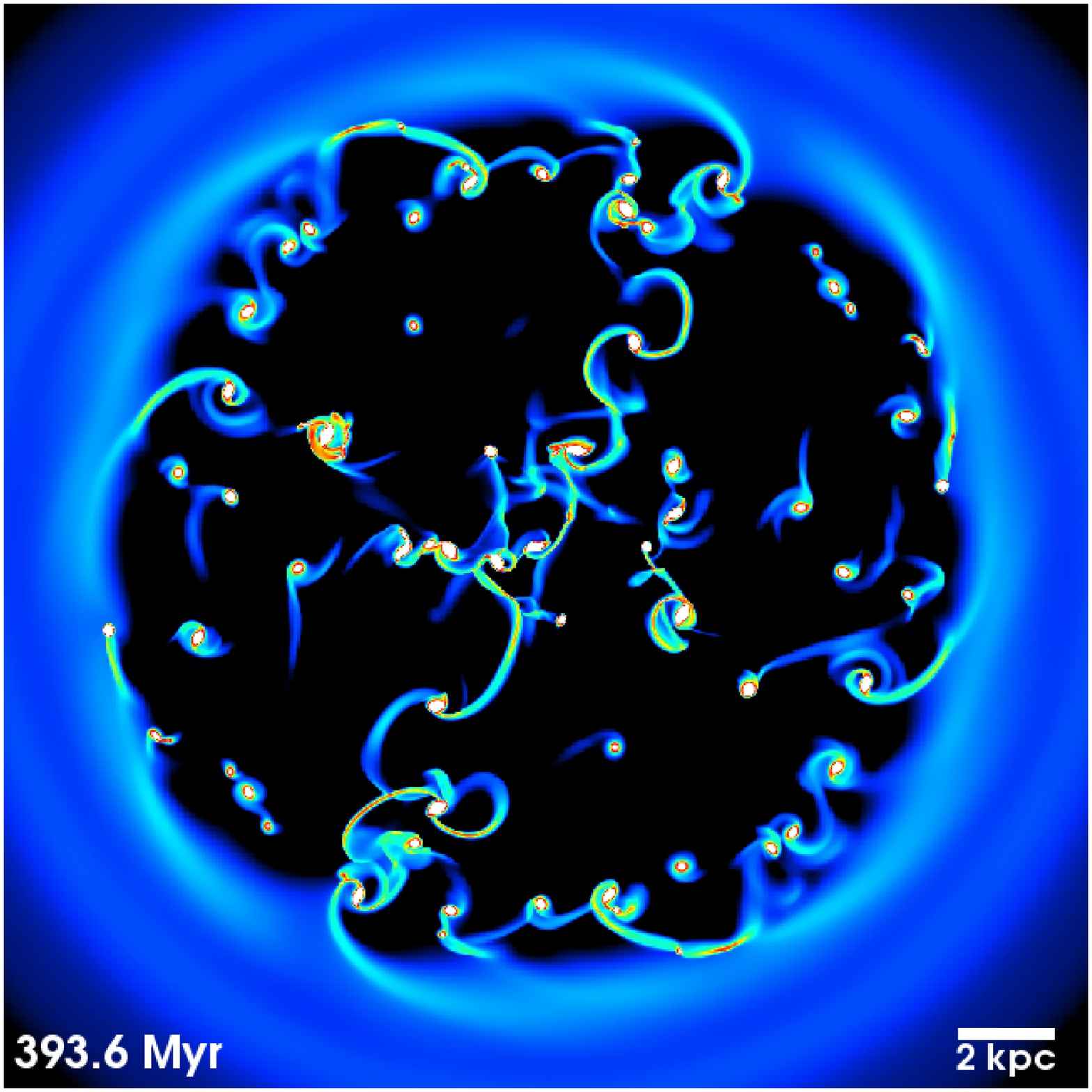}}\hfill
\subfloat[ \label{fig:sim15}]
  {\includegraphics[width=.248\linewidth]{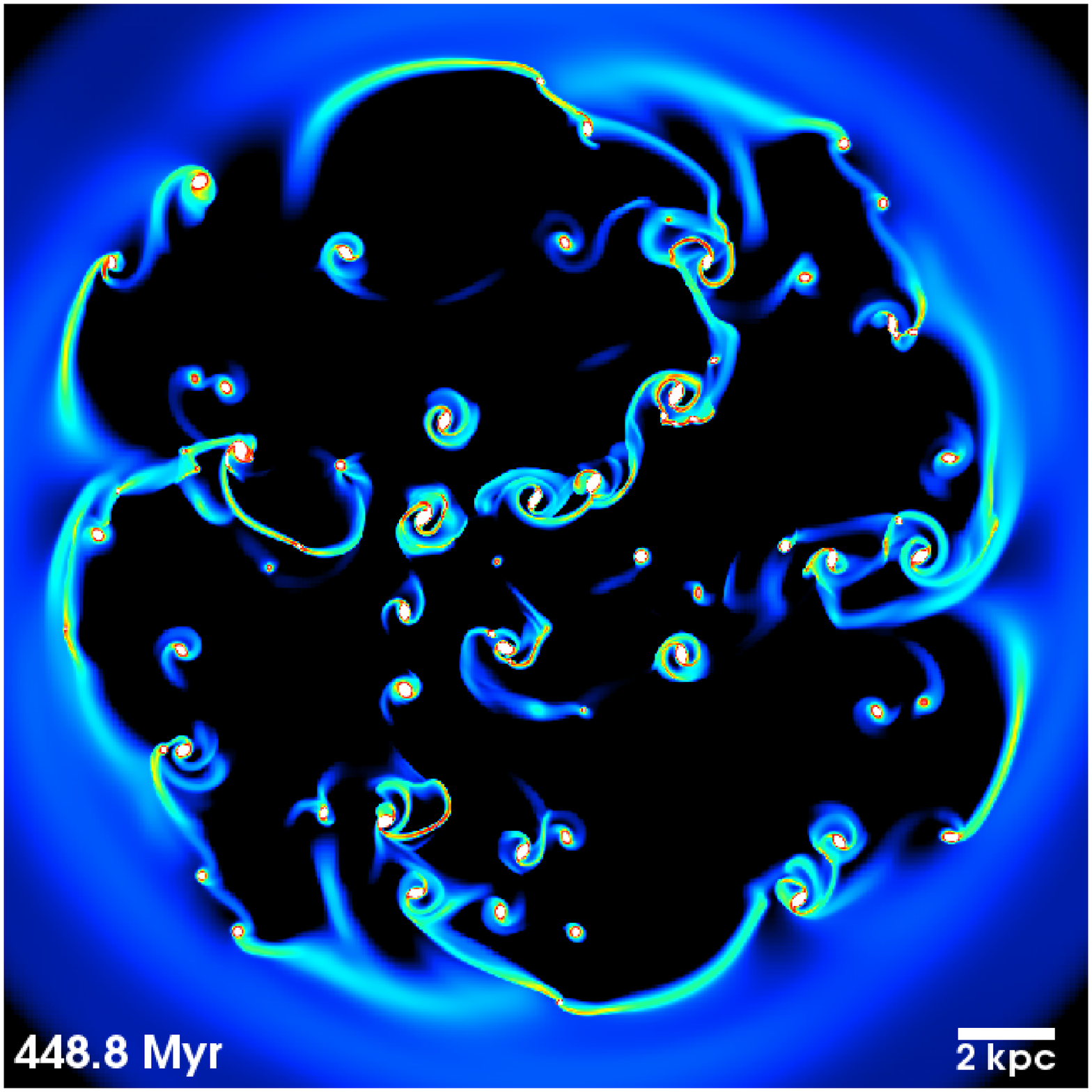}}\hfill
\subfloat[ \label{fig:sim16}]
  {\includegraphics[width=.248\linewidth]{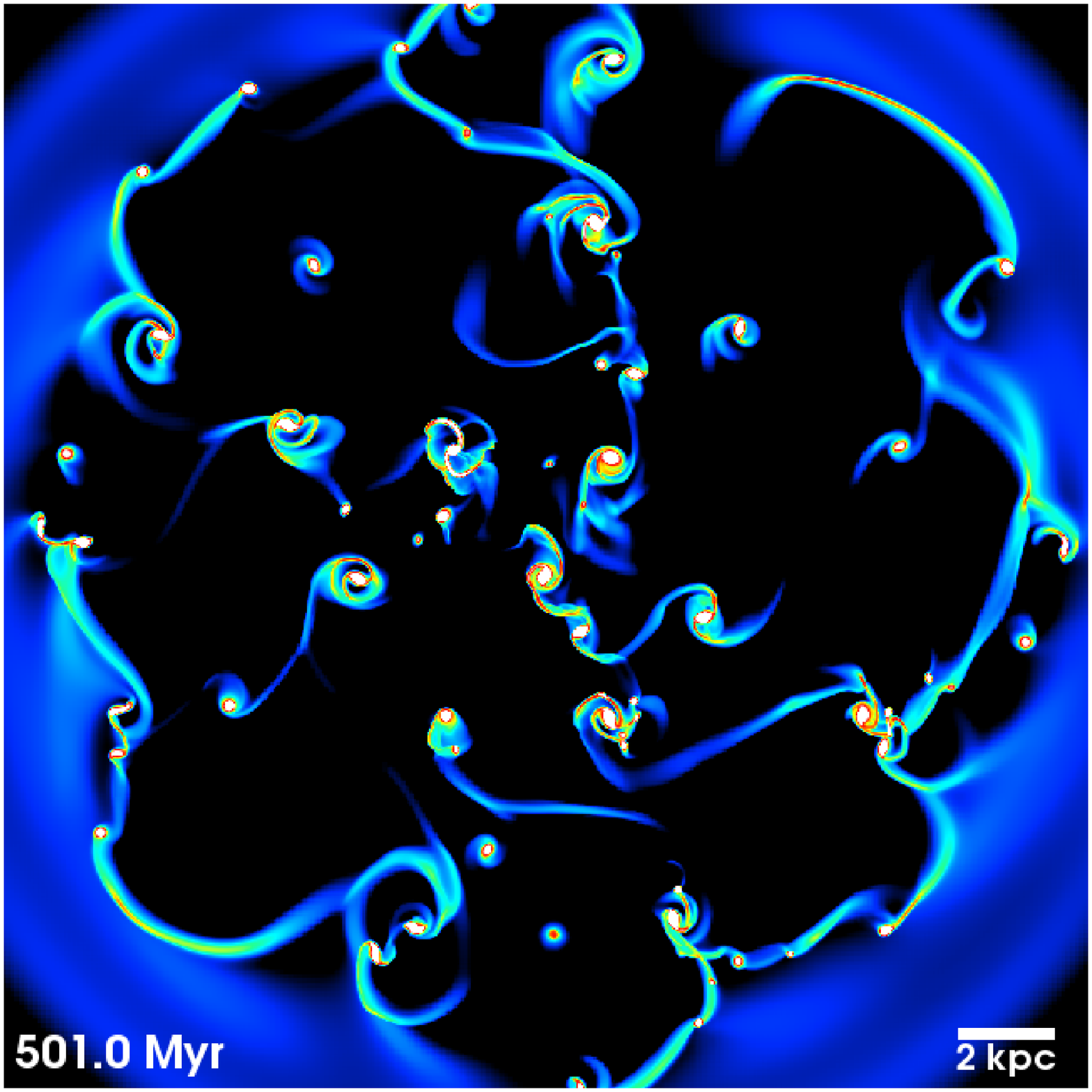}}

\includegraphics[width=.5\linewidth]{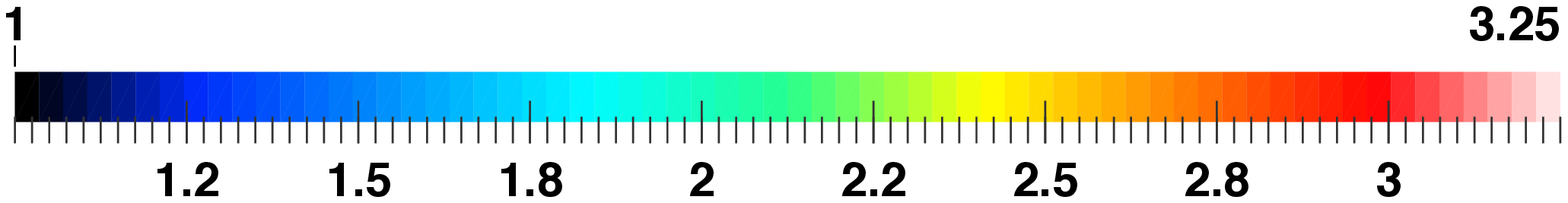} 

\caption{Surface density $\mathrm{log_{10}(\Sigma \; \mathrm{M_{\sun}^{-1} \; pc^2})}$ face-on view of the gas disc simulation at different timesteps. To illustrate the growing rings better, we limit the range of the visible densities. The upper limit is given by $1.8 \times 10^{3} \; \mathrm{M_{\sun} \; pc^{-2}}$, while densities of $\sim 10^5 \; \mathrm{M_{\sun} \; pc^{-2}}$ are reached within some clumps. \label{Results:fig:disk_sim_A}}
\end{figure*}

Axisymmetric overdensities (rings) are forming inside--out (Fig. \ref{Results:fig:disk_sim_A}), as expected from our considerations in Section \ref{Results:subsec:Perturbation theory}. They grow discretely at a certain radius, and their surface density increases, while they accrete mass from the inter-ring regions. At a certain point they begin to collapse to a thin circular line and finally break up into several bound clumps. A highly irregular and clumpy disc is developing within 500 Myr. \\
The first structures become visible in the surface density after $50$ Myr (Figs \ref{Results:fig:disk_sim_A} a and b) and evolve to distinct rings (Figs \ref{Results:fig:disk_sim_A}c-h). At later times, outer rings form with larger radial widths and on longer time-scale (Figs \ref{Results:fig:disk_sim_A}i-p). The last visible ring cannot fully evolve, since it is getting disturbed by the inner clumpy structure (Figs \ref{Results:fig:disk_sim_A}k-p).

\subsection{Perturbation growth}
\label{Results:subsec:perturbation growth}
\begin{figure}
\includegraphics[width=84mm]{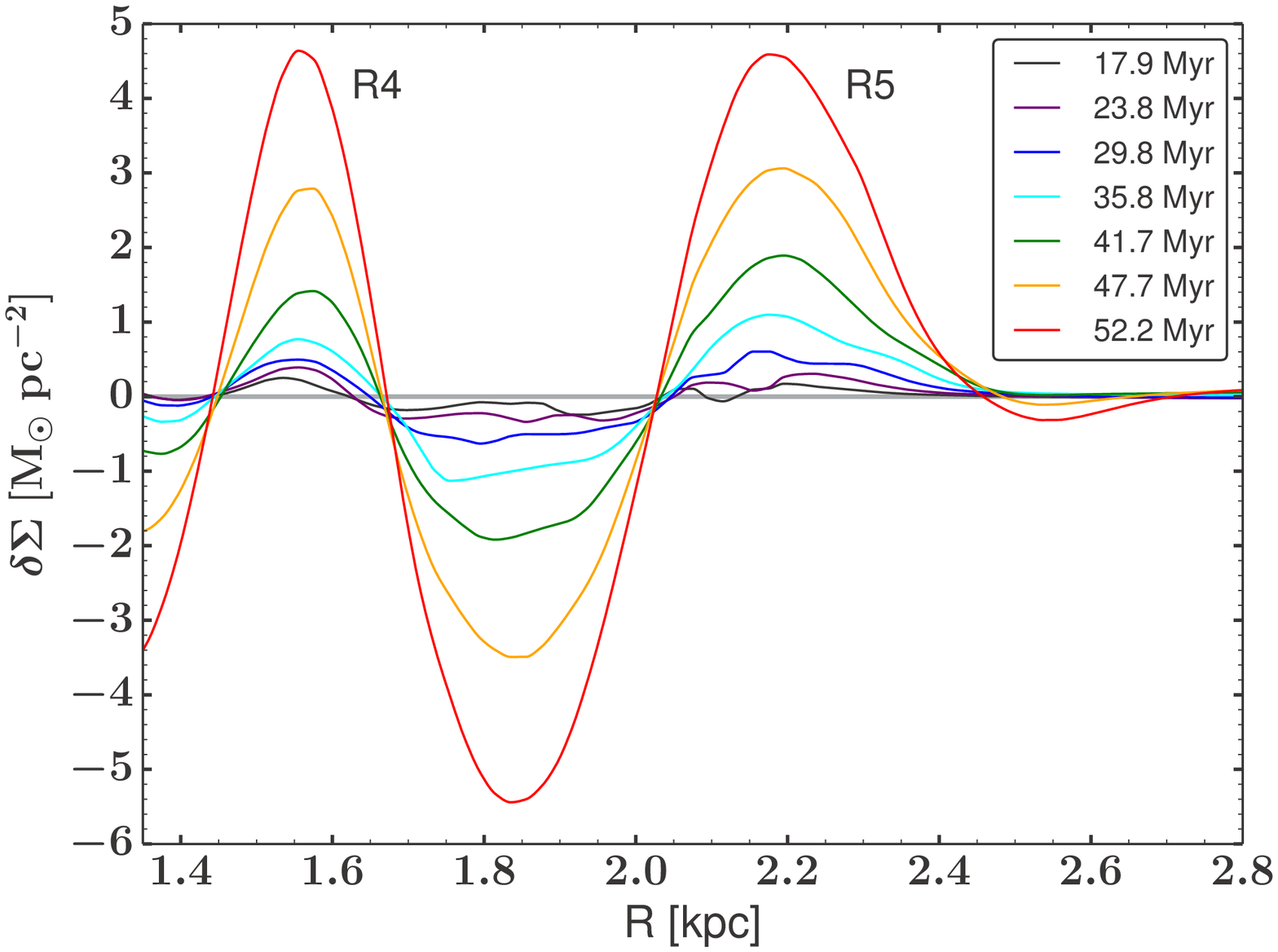}
\caption{$\delta \Sigma$ profiles of rings $R4$ and $R5$ as function of time, to illustrate the clear growth of the ring structure and its wave shape character. At earlier times, superimposed fluctuations with different wavelengths and similar amplitudes appear, which makes it difficult to quantify the inflection points of the dominant wave (see Fig. \ref{Results:fig:bubble_plot}). Later, one dominant wave establishes itself from the spectrum, which can be already identified at very early times.
\label{U_Zoom_Rings_2}} 
\end{figure}

\begin{figure*}
  \centering 
\includegraphics[width=150mm]{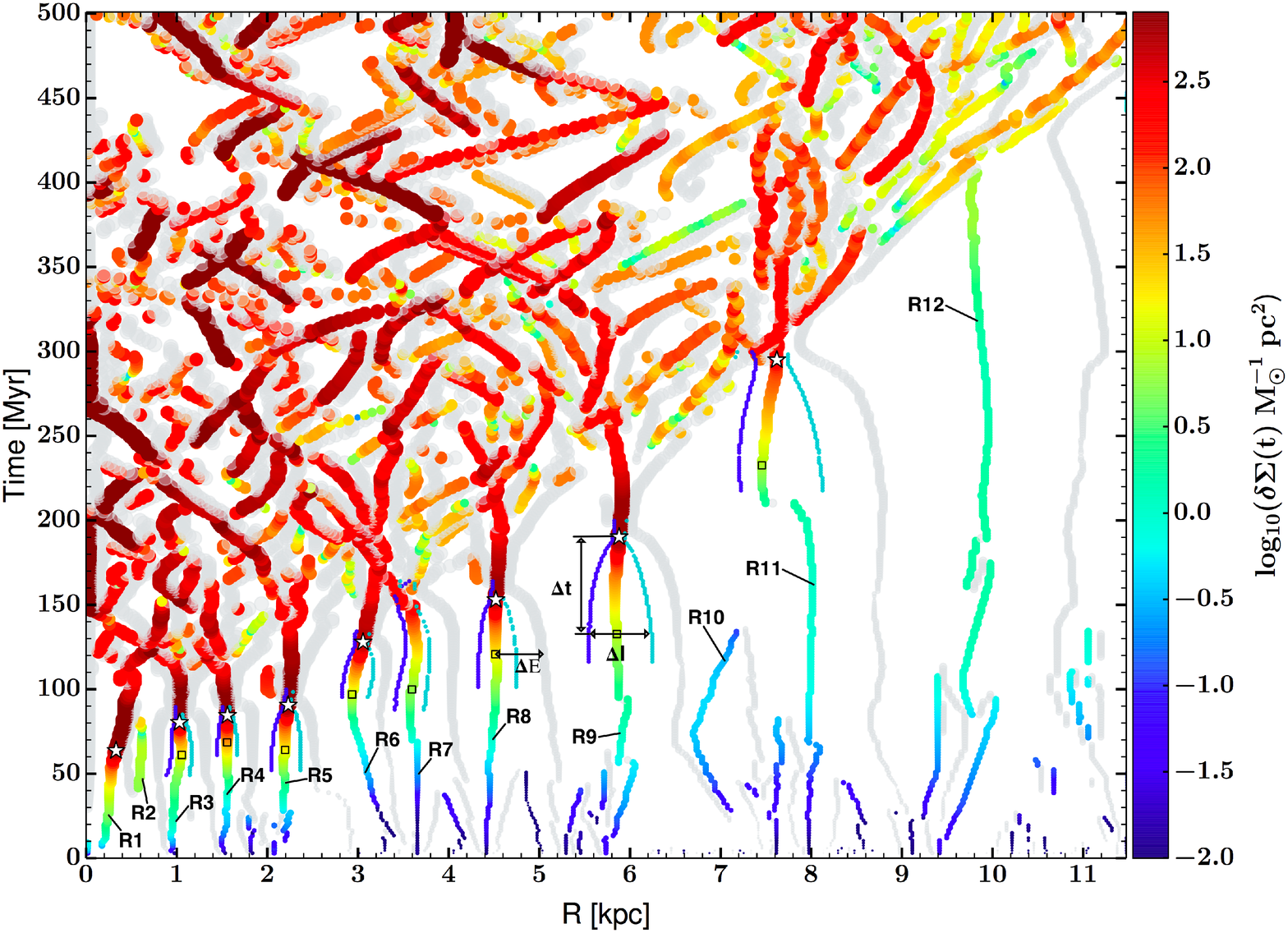} 
\caption{Growth of the overdensities $\delta \Sigma (t)$ (relative extrema) over time and radius (see Section \ref{Results:subsec:perturbation growth}). The colour and the size of the circles illustrate the amplitude of the relative maxima, whereas the grey circles represent the minima. To quantify the size of the fastest growing perturbation wavelength, we use the distance $\Delta I$ between the inflection points (dark-blue and cyan lines), corresponding to the overdense region of the ring $L_{\mathrm{sech^2}}$ (see Fig. \ref{Results:fig:ring_sizes}). It is difficult to determine the growing perturbations in the initial fluctuation spectrum (see Fig. \ref{U_Zoom_Rings_2}). We therefore specify the location of the inflection points when they have clearly formed ($t_{\mathrm{g}}$), that is shortly before they begin to collapse (open squares). The star symbols mark the time $t_{\bigstar}$ when the density perturbations have gravitationally collapsed, shortly before they break up into clumps. The collapse time-scale $\Delta t$ is the difference between $t_{\mathrm{g}}$ and $t_{\bigstar}$. $\Delta E$ is the distance between the maxima and their minima to the right at $t_{\mathrm{g}}$. The clumpy disc is represented by the chaotic upper part of the plot.
\label{Results:fig:bubble_plot}} 
\end{figure*}

\begin{figure*} 
\centering
\subfloat[  \label{fig:linear_fit}]
  {\includegraphics[width=.48\linewidth]{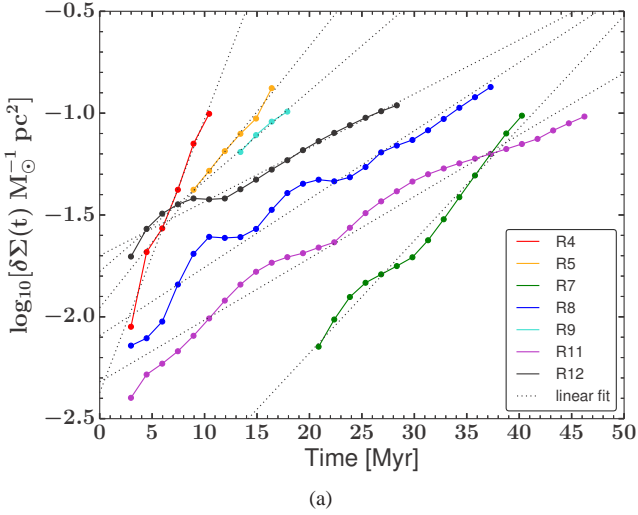}}\hfill
\subfloat[ \label{fig:growth_rates}]
  {\includegraphics[width=.48\linewidth]{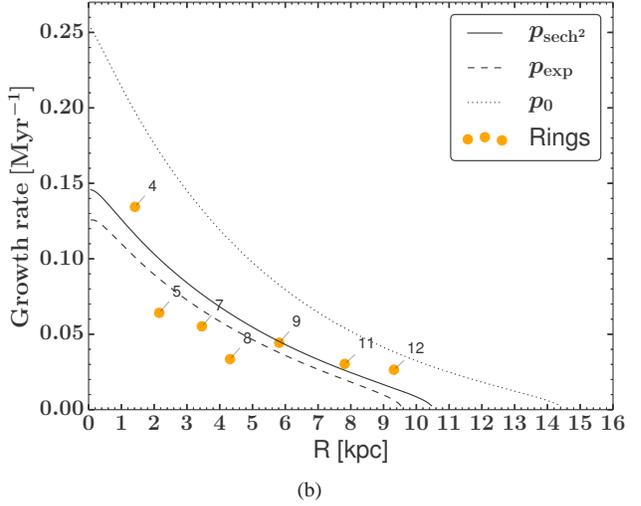}}\hfill
\caption{(a) The logarithm of the ring overdensities over time (coloured lines) in the linear regime within the range  of $\sim 10^{-2.5}-10^{-1} \mathrm{M_{\sun} \ pc^{-2}}$. We make linear fits (dotted lines) to the logarithm of these amplitudes and consider only the fluctuations for which we have at least four snapshots. (b) The radius dependent growth rates $p = \sqrt{- \omega^2}$ (Section \ref{subsec:Timescales}) correspond to the slopes of the linear fits in Fig. \ref{fig:linear_fit} (orange circles). The black lines represent the theoretically derived growth rates of the different vertical density profiles and the infinitesimally thin disc with $p_0$. }
\end{figure*}

The appearing rings are a result of growing perturbations of the dominant modes, from a very small density perturbation seed, caused by a superimposed wave spectrum, due to the initial AMR grid. To follow this process, we determine the relative changes of the surface density compared to the initial disc for several timesteps. Due to axisymmetry, we only consider one quadrant of the disc and average the surface density azimuthally within $\Delta R =20$ pc bins. At each time step ($\sim$1.49 Myr), we subtract the initial surface density profile from the current one $(\delta \Sigma = \Sigma_t - \Sigma_0)$. This method is only possible, because of the, initially, relatively well-balanced disc setup, and gives the quantities, which arise from the perturbation growth with a relatively clear measurable signal (see Fig. \ref{U_Zoom_Rings_2}). We measure the initial seed overdensities with $\delta \Sigma (t=0)  \sim 10^{-3} \; \mathrm{M_{\sun} \;pc^{-2}}$. They grow up to maximum values $\delta \Sigma (t) \sim 10^5 \; \mathrm{M_{\sun} \; pc^{-2}}$ over time within the clumps. In Fig. \ref{Results:fig:bubble_plot}, we limit the range to $\delta \Sigma =10^{-2} - 8 \times 10^2 \; \mathrm{M_{\sun} \;pc^{-2}}$ to illustrate the amplitude of the ring overdensities. Most of the low-density disturbances disappear very early in the evolution. Only a few survive and lead to very high densities, form rings and finally fragment into a clumpy, disordered structure. The innermost ring, $R1$, is moving slightly outwards and comes very near to the second one, $R2$, while it is growing and finally begins to absorb $R2$ completely before it can break up into fragments. The rings $R3$, $R4$, $R5$, $R8$, $R9$ remain well isolated during their evolution. $R7$ is moving towards $R6$, while it is fragmenting (see also Fig. \ref{Results:fig:disk_sim_A} (e)-(g)).  $R10$ and $R12$  are not fully developing. While $R12$ is disrupted by the clumpy structure further in, before its fragmentation, $R10$ is moving outwards and is merging into a new maximum together with $R11$, which is causing the shift inwards of $R11$ at roughly $t=200$ Myr. During the merging, $R10$ is not detectable as a maximum because of its small amplitude.\\
In the beginning, the growth is exponential and therefore the rate is constant. This phase corresponds typically to overdensities of $\sim 10^{-2.5}-10^{-1} \mathrm{M_{\sun} \ pc^{-2}}$, which we define as the linear domain. We make linear fits to the logarithm of these amplitudes over time and consider only the rings for which we have at least four snapshots (Fig. \ref{fig:linear_fit}). The measured slopes are in good agreement with the corresponding radius dependent theoretical growth rates (Section \ref{subsec:Timescales}) $p_{\mathrm{sech^2}} = \sqrt{- \omega^2_{\mathrm{sech}^2}}$ and $p_{\mathrm{exp}}$ (Fig. \ref{fig:growth_rates}), while $p_0$ for an infinitesimally thin disc deviates strongly as expected.

\subsection{Ring properties}
\label{Results:subsec:Ring properties}

\begin{figure}
\includegraphics[width=84mm]{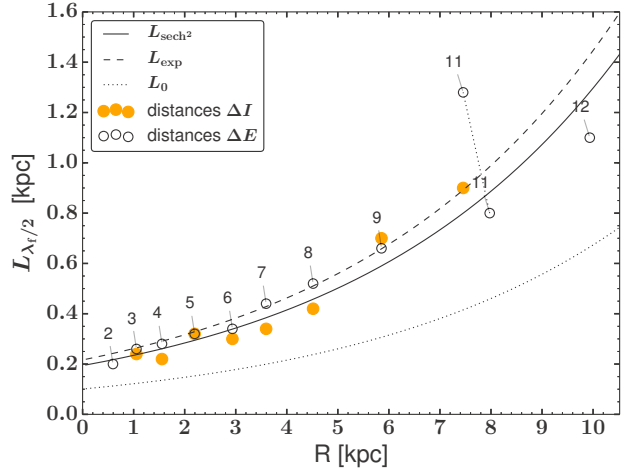} 
\caption{Theoretically derived radial thickness of the rings, which is half the fastest growing wavelength (see Section \ref{subsec:Ring properties}) at every radius (lines). $L_{\mathrm{sech^2}}$ and $L_{\mathrm{0}}$ (equation \ref{eq:clump_radius_S}) correspond to the vertical $\sech^2$ density profile and to the thin disc approximation, respectively. For the vertical exponential profile approximation $L_{\mathrm{exp}} = \lambda_{\mathrm{f_{exp}}} / 2$, we use equation \ref{eq:fastest_growing_wavelength_exp}. The data points are measured in the simulation at times $t_{\mathrm{g}}$, when the rings begin to collapse. The orange circles correspond to the distances between the inflection points $\Delta I$ of the rings, the open circles give the distances between maxima and the minima $\Delta E$, of the radial density distribution for one perturbation wavelength. For the rings $R2$ and $R12$, $t_{\mathrm{g}}$ is not well defined, hence we measure $\Delta E$ at $t=50.7$ and $280.3$ Myr, respectively. The density maximum of ring $R11$ is moving during its evolution (between $t \sim 200$ and $225 $ Myr) and the positions for both times are given by the open circles. Its minimum is not moving, which is causing the huge deviation from the expected value.
\label{Results:fig:ring_sizes}}
\end{figure}

\begin{figure}
\includegraphics[width=84mm]{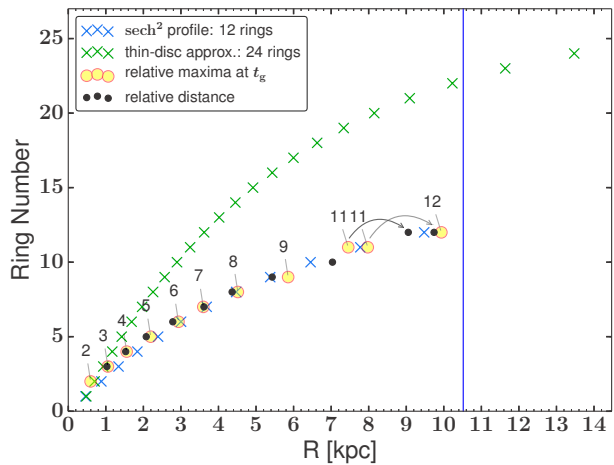} 
\caption{Comparison of the measured ring positions and numbers with the theoretical predictions within their unstable region. Here, we assume that the local fastest growing perturbation wavelength gives the distance to the next growing ring from inside--out. The blue crosses correspond to the $\sech^2$ density profile and the green to the thin disc approximation. The orange data points show the measured positions of the relative density maximum at the time $t_{\mathrm{g}}$ (see also explanation in Fig. \ref{Results:fig:ring_sizes}). For the black data points, we calculate the local perturbation wavelength relative for each orange data point to get every relative distance ($\sech^2$ profile). The maxima of ring $R11$ is moving slightly inwards, therefore two data points are given.
\label{Results:fig:ring_positions}}
\end{figure}

To quantify the radial thickness of the rings, which is half of the fastest growing wavelength, we measure the distances between the inflection points $\Delta I$ of the perturbation (see Fig. \ref{Results:fig:bubble_plot}). In the beginning, it is difficult to determine them from the low amplitude superimposed wavelength spectrum (e.g. Fig. \ref{U_Zoom_Rings_2}). For the very low amplitudes the interactions of many initial perturbations, including those that do not grow, lead to several inflection points. With time the amplitudes of the growing wavelengths begin to dominate. At that point their inflection points can be measured properly. Therefore, we consider for the inflection points only the clearly evolved rings, a few timesteps before they begin to collapse due to self-gravity at time $t_{\mathrm{g}}$. There we reach the time, when the rings' self-gravity is strong enough to begin the collapse. The minimum distance of the inflection points we define as the time $t_{\bigstar}$, just before the rings break up into clumps (marked with the symbol $\bigstar$ in Fig. \ref{Results:fig:bubble_plot}). Additionally, as a second constraint for the wavelength, we measure the distance $\Delta E$ between the relative maxima and minima (to the right, shown by the grey circles in Fig. \ref{Results:fig:bubble_plot}) at time $t_{\mathrm{g}}$. The radial sizes $\Delta I$ and $\Delta E$, measured at $t_{\mathrm{g}}$ (see Fig. \ref{Results:fig:ring_sizes}), are in good agreement with the theoretical expectations for the $\sech^2$ profile (see Section \ref{subsec:Ring properties}). The relative maximum of $R11$ is moving slightly inwards but not its already strongly evolved minimum, which causes the deviation from the expectation. The thin disc approximation assumes too small and the exponential density approximation slightly too large structures. At time $t_{\mathrm{g}}$ we can also determine the positions of eight rings within the disc (Fig. \ref{Results:fig:ring_positions}) and compare them to the theory in two ways. First, we take the positions of every measured ring and add radially to each the theoretical local perturbation wavelengths $\lambda_{f_{\mathrm{sech^2}}}$. This indeed gives us the opportunity to infer the distances to each of the next growing ring position, within small deviations. Most of the rings can be explained by this method, only Ring $R9$ has a larger deviation from the expectation. Furthermore, we also expect an additional ring $R10$, at $R \sim 7$ kpc, which is seen as an early phase in the evolution with its corresponding minima (see Fig. \ref{Results:fig:bubble_plot}), but which merges with $R11$. 

Alternatively, we start at a point near to the disc centre, take only the theoretically calculated local perturbation wavelength $\lambda_{f_{\mathrm{sech^2}}}$ and add it to infer the next position from where we repeat the process until the end of the unstable regime. All the following theoretical ring positions are dependent on the first ring. This is shown in Fig. \ref{Results:fig:ring_positions}. We choose the theoretical position of $R1$ in order to match most of the growing rings in the simulation. In our selection we are in quite good agreement with the measured rings, again an additional ring $R10$ is expected. If we repeat the same by using the perturbation wavelength $\lambda_{f_{\mathrm{exp}}}$ calculated in its unstable regime, we would expect a maximum of 10 rings and their predicted positions only partly agree with the simulation. With $\lambda_{f_0}$, we would expect 24 rings for the unstable regime of the thin disc approximation which is very different from the result of our simulation.

\subsection{Time-scales}
\label{Results:subsec:Timescales}
\begin{figure}
\includegraphics[width=84mm]{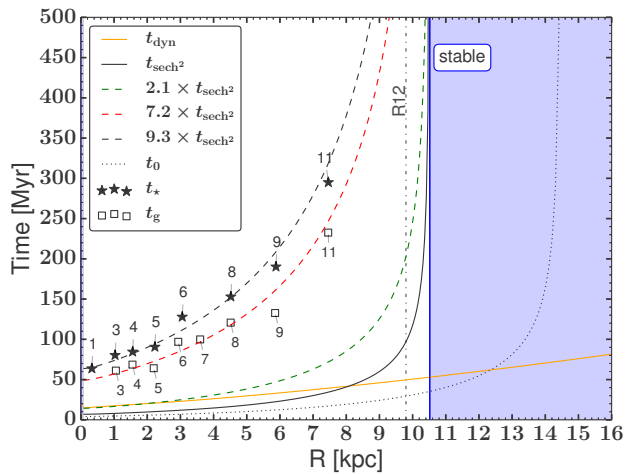}
\caption{Demonstration of the different time-scales (see Section \ref{subsec:Timescales} and Fig. \ref{Results:fig:bubble_plot}). The solid black line gives the theoretical time-scale $t_{\mathrm{sech^2}}$, in which the perturbations can grow by a factor of $e$. The open squares give the time $t_{\mathrm{g}}$ when the ring-like structures begin to collapse, the star symbols the time $t_{\mathrm{\bigstar}}$, when they reach the minimum thickness, shortly before breaking up into clumps. The $t_{\mathrm{g}}$ measurements follow $7.2 \times t_{\mathrm{sech^2}}$ (red dashed line) and the data for $t_{\mathrm{\bigstar}}$  is close to $9.3 \times t_{\mathrm{sech^2}}$ (black 
ed line). The resulting collapsing time-scale $\Delta t$ between these two time-scales is $2.1 \times t_{\mathrm{sech^2}}$ (green dashed line) and is very similar to the dynamical crossing time $t_{\mathrm{dyn}}$ in the inner region ($R \leq 3.5$ kpc). The dash--dotted line gives roughly the position of $R12$ and shows that the expected theoretical fragmentation is at very late times, compared to the ring structures that develop inside the disc. The thin disc approximation gives a far shorter time-scale (dotted line), and predicts fragmentation even in the already stable region according to linear analysis (blue shaded).
\label{Results:fig:timescales}} 
\end{figure}

Here we compare our disc simulation to the time-scale $t_{\mathrm{sech^2}}$ (\ref{subsec:Timescales}), when the perturbation density $\Sigma$ has increased by a factor of $e$ (Fig. \ref{Results:fig:timescales}). The growth time remains rather short up to $\sim 6$ kpc and finally rises strongly and goes to infinity at $Q_{\mathrm{sech^2}}=1$. If we rewrite equation (\ref{eq:t_sech^2}) by inserting $Q_{0}$, we get
\begin{equation}
\label{eq:Results:t_sech^2}
t_{\mathrm{sech^2}} = \left(  D \;  \Sigma^2 - \kappa^2  \right)^{-1/2},
\end{equation}
with the constant factor $D = \left( \frac{\upi G}{C_{\mathrm{sech^2}} \; c_{\mathrm{s}}} \right)^2$ and the only two radial dependent parameters $\Sigma$ and $\kappa$. The surface density, which plays a destabilizing role, and the epicyclic frequency which stabilizes, decrease both with radius, but $\Sigma$ decreases faster, especially for the last third of the unstable regime. The shape of the time-scale dependence on radius stays similar from the beginning of the growing structures up to the collapsed rings. This means that the growth time can roughly be expressed by a constant factor $k$ times $t_{\mathrm{sech^2}}$ at every radius ($k \times t_{\mathrm{sech^2}}$).\\
The rings begin to collapse at $t_{\mathrm{g}} \approx 7.2 \times t_{\mathrm{sech^2}}$ (dashed red line), and reach maximum density at $t_{\bigstar} \approx 9.3 \times t_{\mathrm{sech^2}}$ (dashed black line), when the simulation attains the resolution limit ($t_{\mathrm{g}}$ and $t_{\bigstar}$ are the arithmetic mean of the corresponding data points). The collapsing time-scale $\Delta t$ therefore is $2.1 \times t_{\mathrm{sech^2}}$, which is very similar to the dynamical crossing time $t_{\mathrm{dyn}} = \frac{R}{V_{\mathrm{rot}}}$ for the inner $R = 3.5$ kpc, but is much larger with increasing radius. For the time-scale $t_0$ of the thin disc approximation we could not find a factor to describe the measured times $t_{\mathrm{g}}$ and $t_{\mathrm{\bigstar}}$.

\section{Discussion and Conclusions}
\label{sec:Discussion and Conclusions}
We studied the structure formation due to gravitational instability in a self-gravitating gas disc in greater detail. The axisymmetric perturbation theory is revisited by taking its finite thickness with a typical vertical $\sech^2$ density profile into account. To test the derived properties, we employed idealized simulations of an isothermal gas disc in hydrodynamical equilibrium, unstable to axisymmetric perturbations. From the linear stability analysis follows:
\begin{enumerate}
\item  In the unstable regime, the fastest growing perturbation wavelength $\lambda_{f_{\mathrm{sech^2}}} $ of a vertical $\sech^2$ density profile is always $1.926$ times larger than in the classical razor-thin disc approximation $\lambda_{f_{0}} = \frac{2 c_{\mathrm{s}}^2}{G \Sigma}$ and differs therefore by $48.079 \%$. The widely used approximation of an exponential profile leads to a wavelength $\lambda_{f_{\mathrm{exp}}}$ which is always $2.148$ times larger than $\lambda_{f_{0}}$ and gives an error of $\sim 11.527 \%$ compared to $\lambda_{f_{\mathrm{sech^2}}} $. These ratios are independent of the disc scaleheight and therefore the temperature or surface density of the disc and lead to a simple correlation between $\lambda_{\mathrm{f_{sech^2}}}$ and $\lambda_{f_{\mathrm{0}}}$ in hydrostatic equilibrium. 
\item In contrast to other derivations we apply the fastest growing wavelength on the dispersion relation to determine the Toomre instability parameter and find for thick discs with a $\sech^2$ profile the critical value $Q_{\mathrm{0, crit}} \simeq 0.696$ which is very similar to the value found by \citet{Wang:2010wr} with $Q_{\mathrm{0,Wang}} \simeq 0.693$. For the exponential density approximation, we found the same value as in \citet{Kim:2002gl} with $Q_{\mathrm{0, KOS}} \simeq 0.647$. Also here, the relations to the razor-thin disc instability parameter are independent of the disc scaleheight. The classical Toomre $Q$ parameter overestimates local self-gravity and leads to the assumption of a too large radial unstable regime, while the exponential approximation underestimates local self-gravity and infers a slightly too small unstable region in our disc model.
\end{enumerate}
In order to test the analytical solution and to explore the transition into the highly non-linear regime, we compare it with our hydrodynamical simulations and can conclude.
\begin{enumerate}
\item In the first phase, rings form that organize themselves discretely, with distances corresponding to the local fastest growing perturbation wavelength $\lambda_{f_{\mathrm{sech^2}}}(R)$. The radial thickness of the measured overdensity is compatible to half of the wavelength $L_{\mathrm{sech^2}} (R)= \lambda_{f_{\mathrm{sech^2}}}(R) / 2$. The total number of growing rings, calculated in their respective unstable regime, is for the exponential profile approximation underestimated by $\sim17 \%$ and their predicted positions only partly agree with the simulation. For the thin disc approach, the number is overestimated by a factor of 2. \\
For an isothermal disc, both features, ring size (radial thickness) and distance, increase with radius where also the time-scales increase steeply when approaching the stable regime. While the perturbation wavelength is only dependent on sound speed and surface density (equation \ref{eq:fastest_growing_wavelength_sech2}), the stability and growth time is additionally dependent on the differential rotation, as $\kappa$ is proportional to the angular frequency $\Omega$ (equation \ref{eq:Results:t_sech^2}). The initial perturbation spectrum is seeded by the AMR grid; however, the fastest growing modes agree with the linear analysis and later on dominate.
\item The individual rings grow in density, and later on contract to thin and dense circular lines, over the same time accreting more gas from the inter-ring regions.
In the beginning, the growth is exponential and therefore the rate is constant up to overdensities of $\sim 0.1 \ \mathrm{M_{\sun} \ pc^{-2}}$, which we define as the linear domain, and which is in good agreement with the linear analysis. The ring growth rates are roughly self-similar, which is reflected in the growth by a constant factor $k$ times $t_{\mathrm{sech^2}}$ at every radius ($\sim k \times t_{\mathrm{sech^2}}$), where $t_{\mathrm{sech^2}}$ is the theoretical time when the perturbation density $\Sigma$ has increased by a factor of $e$. We determine the collapse time-scale $\Delta t$ to be $2.1 \times t_{\mathrm{sech^2}}$, which is similar to the dynamical crossing time within the inner $3.5$ kpc and strongly deviates for larger radii. We  estimate the ring mass to be $M_{\mathrm{Ring}} = 2 \upi \Sigma (R) \; R \; \lambda_{\mathrm{f_{sech^2}}}$, where $R$ is the location of the maximum of the density perturbation. In the thin disc approximation, the mass is underestimated by $\sim 48 \%$ due to the linear proportionality of the mass to the fastest growing wavelength.
\item The dense and thin circular, ring-like filaments finally fragment into a large number of clumps. The clump sizes are not dependent on the initial radial ring widths anymore and cannot be directly predicted from the simple perturbation theory, as is usually assumed. They break up into individual clumps in an evolutionary phase where the rings have already developed strongly non-linear, perturbed regions. At this point, the role of axisymmetric perturbations, the resolution limit, the artificial pressure floor and other physical processes become important. We will investigate the emergence of clumps by ring instabilities in an upcoming publication. Furthermore, non-axisymmetric modes have to be studied in greater detail.
\item In order to guarantee proper growth of the initial ring structures, we find that simulations have to resolve the initial Jeans length in the disc mid-plane with more than $18$ grid cells, which corresponds to about five cells per disc scaleheight.
\end{enumerate}

\section*{Acknowledgements} 
We thank our referee Richard Durisen, for important comments that helped to clarify the paper. We are grateful to Alessandro Romeo and Bruce Elmegreen for comments and Alessandro Ballone for useful discussions. Computer resources for this project have been provided by the Leibniz Supercomputer Centre under grant: h0075. MS was supported by the Deutsche Forschungsgemeinschaft under the priority programme 1573 ('Physics of the Interstellar Medium').

\bibliographystyle{mn2e}
\bibliography{mystrings,mybib}

\begin{thebibliography}{35}
\expandafter\ifx\csname natexlab\endcsname\relax\def\natexlab#1{#1}\fi

\bibitem[{{Agertz}, {Teyssier} \& {Moore}(2009){Agertz}, {Teyssier}, \&
  {Moore}}]{Agertz:2009wd}
{Agertz} O., {Teyssier} R., {Moore} B., 2009, MNRAS, 397, L64

\bibitem[{{Binney} \& {Tremaine}(2008)}]{Binney:2011vb}
{Binney} J., {Tremaine} S., 2008, {Galactic Dynamics}. {Princeton Univ. Press,
  Princeton, NJ}

\bibitem[{{Bournaud}, {Elmegreen} \& {Elmegreen}(2007){Bournaud}, {Elmegreen},
  \& {Elmegreen}}]{2007ApJ...670..237B}
{Bournaud} F., {Elmegreen} B.~G., {Elmegreen} D.~M., 2007, ApJ, 670, 237

\bibitem[{{Bournaud} {et~al}\mbox{.}(2010){Bournaud}, {Elmegreen}, {Teyssier},
  {Block}, \& {Puerari}}]{2010MNRAS.409.1088B}
{Bournaud} F., {Elmegreen} B.~G., {Teyssier} R., {Block} D.~L., {Puerari} I.,
  2010, MNRAS, 409, 1088

\bibitem[{{Bournaud} {et~al}\mbox{.}(2014){Bournaud}, {Perret}, {Renaud},
  {Dekel}, {Elmegreen}, {Elmegreen}, {Teyssier}, {Amram}, {Daddi}, {Duc},
  {Elbaz}, {Epinat}, {Gabor}, {Juneau}, {Kraljic}, \& {Le
  Floch'}}]{2014ApJ...780...57B}
{Bournaud} F. {et~al.}, 2014, ApJ, 780, 57

\bibitem[{{Burkert}(1995)}]{Burkert:1995jr}
{Burkert} A., 1995, ApJ, 447, L25

\bibitem[{{Ceverino}, {Dekel} \& {Bournaud}(2010){Ceverino}, {Dekel}, \&
  {Bournaud}}]{Ceverino:2010eh}
{Ceverino} D., {Dekel} A., {Bournaud} F., 2010, MNRAS, 404, 2151

\bibitem[{{Ceverino} {et~al}\mbox{.}(2012){Ceverino}, {Dekel}, {Mandelker},
  {Bournaud}, {Burkert}, {Genzel}, \& {Primack}}]{2012MNRAS.420.3490C}
{Ceverino} D., {Dekel} A., {Mandelker} N., {Bournaud} F., {Burkert} A.,
  {Genzel} R., {Primack} J., 2012, MNRAS, 420, 3490

\bibitem[{{Daddi} {et~al}\mbox{.}(2010){Daddi}, {Bournaud}, {Walter},
  {Dannerbauer}, {Carilli}, {Dickinson}, {Elbaz}, {Morrison}, {Riechers},
  {Onodera}, {Salmi}, {Krips}, \& {Stern}}]{2010ApJ...713..686D}
{Daddi} E. {et~al.}, 2010, ApJ, 713, 686

\bibitem[{{Dekel}, {Sari} \& {Ceverino}(2009){Dekel}, {Sari}, \&
  {Ceverino}}]{Dekel:2009bn}
{Dekel} A., {Sari} R., {Ceverino} D., 2009, ApJ, 703, 785

\bibitem[{{Elmegreen}(1987)}]{1987ApJ...312..626E}
{Elmegreen} B.~G., 1987, ApJ, 312, 626

\bibitem[{{Elmegreen}(2011)}]{2011ApJ...737...10E}
{Elmegreen} B.~G., 2011, ApJ, 737, 10

\bibitem[{{Elmegreen} \& {Elmegreen}(1983)}]{1983ApJ...267...31E}
{Elmegreen} B.~G., {Elmegreen} D.~M., 1983, ApJ, 267, 31

\bibitem[{{Elmegreen} {et~al}\mbox{.}(2007){Elmegreen}, {Elmegreen},
  {Ravindranath}, \& {Coe}}]{2007ApJ...658..763E}
{Elmegreen} D.~M., {Elmegreen} B.~G., {Ravindranath} S., {Coe} D.~A., 2007,
  ApJ, 658, 763

\bibitem[{{F{\"o}rster Schreiber} {et~al}\mbox{.}(2009){F{\"o}rster Schreiber},
  {Genzel}, {Bouch{\'e}}, {Cresci}, {Davies}, {Buschkamp}, {Shapiro},
  {Tacconi}, {Hicks}, {Genel}, {Shapley}, {Erb}, {Steidel}, {Lutz},
  {Eisenhauer}, {Gillessen}, {Sternberg}, {Renzini}, {Cimatti}, {Daddi},
  {Kurk}, {Lilly}, {Kong}, {Lehnert}, {Nesvadba}, {Verma}, {McCracken},
  {Arimoto}, {Mignoli}, \& {Onodera}}]{2009ApJ...706.1364F}
{F{\"o}rster Schreiber} N.~M. {et~al.}, 2009, ApJ, 706, 1364

\bibitem[{{Gammie}(2001)}]{Gammie:2001hv}
{Gammie} C.~F., 2001, ApJ, 553, 174

\bibitem[{{Genzel} {et~al}\mbox{.}(2008){Genzel}, {Burkert}, {Bouch{\'e}},
  {Cresci}, {F{\"o}rster Schreiber}, {Shapley}, {Shapiro}, {Tacconi},
  {Buschkamp}, {Cimatti}, {Daddi}, {Davies}, {Eisenhauer}, {Erb}, {Genel},
  {Gerhard}, {Hicks}, {Lutz}, {Naab}, {Ott}, {Rabien}, {Renzini}, {Steidel},
  {Sternberg}, \& {Lilly}}]{2008ApJ...687...59G}
{Genzel} R. {et~al.}, 2008, ApJ, 687, 59

\bibitem[{{Genzel} {et~al}\mbox{.}(2011){Genzel}, {Newman}, {Jones},
  {F{\"o}rster Schreiber}, {Shapiro}, {Genel}, {Lilly}, {Renzini}, {Tacconi},
  {Bouch{\'e}}, {Burkert}, {Cresci}, {Buschkamp}, {Carollo}, {Ceverino},
  {Davies}, {Dekel}, {Eisenhauer}, {Hicks}, {Kurk}, {Lutz}, {Mancini}, {Naab},
  {Peng}, {Sternberg}, {Vergani}, \& {Zamorani}}]{2011ApJ...733..101G}
{Genzel} R. {et~al.}, 2011, ApJ, 733, 101

\bibitem[{{Griv} \& {Gedalin}(2012)}]{2012MNRAS.422..600G}
{Griv} E., {Gedalin} M., 2012, MNRAS, 422, 600

\bibitem[{{Grogin} {et~al}\mbox{.}(2011){Grogin}, {Kocevski}, {Faber},
  {Ferguson}, {Koekemoer}, {Riess}, {Acquaviva}, {Alexander}, {Almaini},
  {Ashby}, {Barden}, {Bell}, {Bournaud}, {Brown}, {Caputi}, {Casertano},
  {Cassata}, {Castellano}, {Challis}, {Chary}, {Cheung}, {Cirasuolo},
  {Conselice}, {Roshan Cooray}, {Croton}, {Daddi}, {Dahlen}, {Dav{\'e}}, {de
  Mello}, {Dekel}, {Dickinson}, {Dolch}, {Donley}, {Dunlop}, {Dutton}, {Elbaz},
  {Fazio}, {Filippenko}, {Finkelstein}, {Fontana}, {Gardner}, {Garnavich},
  {Gawiser}, {Giavalisco}, {Grazian}, {Guo}, {Hathi}, {H{\"a}ussler},
  {Hopkins}, {Huang}, {Huang}, {Jha}, {Kartaltepe}, {Kirshner}, {Koo}, {Lai},
  {Lee}, {Li}, {Lotz}, {Lucas}, {Madau}, {McCarthy}, {McGrath}, {McIntosh},
  {McLure}, {Mobasher}, {Moustakas}, {Mozena}, {Nandra}, {Newman}, {Niemi},
  {Noeske}, {Papovich}, {Pentericci}, {Pope}, {Primack}, {Rajan},
  {Ravindranath}, {Reddy}, {Renzini}, {Rix}, {Robaina}, {Rodney}, {Rosario},
  {Rosati}, {Salimbeni}, {Scarlata}, {Siana}, {Simard}, {Smidt}, {Somerville},
  {Spinrad}, {Straughn}, {Strolger}, {Telford}, {Teplitz}, {Trump}, {van der
  Wel}, {Villforth}, {Wechsler}, {Weiner}, {Wiklind}, {Wild}, {Wilson},
  {Wuyts}, {Yan}, \& {Yun}}]{2011ApJS..197...35G}
{Grogin} N.~A. {et~al.}, 2011, ApJS, 197, 35

\bibitem[{{Immeli} {et~al}\mbox{.}(2004{\natexlab{a}}){Immeli}, {Samland},
  {Gerhard}, \& {Westera}}]{2004A&A...413..547I}
{Immeli} A., {Samland} M., {Gerhard} O., {Westera} P., 2004{\natexlab{a}},
  A\&A, 413, 547

\bibitem[{{Immeli} {et~al}\mbox{.}(2004{\natexlab{b}}){Immeli}, {Samland},
  {Westera}, \& {Gerhard}}]{2004ApJ...611...20I}
{Immeli} A., {Samland} M., {Westera} P., {Gerhard} O., 2004{\natexlab{b}}, ApJ,
  611, 20

\bibitem[{{Kim}, {Kim} \& {Ostriker}(2006){Kim}, {Kim}, \&
  {Ostriker}}]{2006ApJ...649L..13K}
{Kim} C.-G., {Kim} W.-T., {Ostriker} E.~C., 2006, ApJ, 649, L13

\bibitem[{{Kim} \& {Ostriker}(2007)}]{Kim:2007ek}
{Kim} W.-T., {Ostriker} E.~C., 2007, ApJ, 660, 1232

\bibitem[{{Kim}, {Ostriker} \& {Stone}(2002){Kim}, {Ostriker}, \&
  {Stone}}]{Kim:2002gl}
{Kim} W.-T., {Ostriker} E.~C., {Stone} J.~M., 2002, ApJ, 581, 1080

\bibitem[{{Koekemoer} {et~al}\mbox{.}(2011){Koekemoer}, {Faber}, {Ferguson},
  {Grogin}, {Kocevski}, {Koo}, {Lai}, {Lotz}, {Lucas}, {McGrath}, {Ogaz},
  {Rajan}, {Riess}, {Rodney}, {Strolger}, {Casertano}, {Castellano}, {Dahlen},
  {Dickinson}, {Dolch}, {Fontana}, {Giavalisco}, {Grazian}, {Guo}, {Hathi},
  {Huang}, {van der Wel}, {Yan}, {Acquaviva}, {Alexander}, {Almaini}, {Ashby},
  {Barden}, {Bell}, {Bournaud}, {Brown}, {Caputi}, {Cassata}, {Challis},
  {Chary}, {Cheung}, {Cirasuolo}, {Conselice}, {Roshan Cooray}, {Croton},
  {Daddi}, {Dav{\'e}}, {de Mello}, {de Ravel}, {Dekel}, {Donley}, {Dunlop},
  {Dutton}, {Elbaz}, {Fazio}, {Filippenko}, {Finkelstein}, {Frazer}, {Gardner},
  {Garnavich}, {Gawiser}, {Gruetzbauch}, {Hartley}, {H{\"a}ussler},
  {Herrington}, {Hopkins}, {Huang}, {Jha}, {Johnson}, {Kartaltepe},
  {Khostovan}, {Kirshner}, {Lani}, {Lee}, {Li}, {Madau}, {McCarthy},
  {McIntosh}, {McLure}, {McPartland}, {Mobasher}, {Moreira}, {Mortlock},
  {Moustakas}, {Mozena}, {Nandra}, {Newman}, {Nielsen}, {Niemi}, {Noeske},
  {Papovich}, {Pentericci}, {Pope}, {Primack}, {Ravindranath}, {Reddy},
  {Renzini}, {Rix}, {Robaina}, {Rosario}, {Rosati}, {Salimbeni}, {Scarlata},
  {Siana}, {Simard}, {Smidt}, {Snyder}, {Somerville}, {Spinrad}, {Straughn},
  {Telford}, {Teplitz}, {Trump}, {Vargas}, {Villforth}, {Wagner}, {Wandro},
  {Wechsler}, {Weiner}, {Wiklind}, {Wild}, {Wilson}, {Wuyts}, \&
  {Yun}}]{2011ApJS..197...36K}
{Koekemoer} A.~M. {et~al.}, 2011, ApJS, 197, 36

\bibitem[{{Lin} \& {Shu}(1964)}]{Lin:1964cx}
{Lin} C.~C., {Shu} F.~H., 1964, ApJ, 140, 646

\bibitem[{{Romeo} \& {Agertz}(2014)}]{2014MNRAS.442.1230R}
{Romeo} A.~B., {Agertz} O., 2014, MNRAS, 442, 1230

\bibitem[{{Shetty} \& {Ostriker}(2006)}]{2006ApJ...647..997S}
{Shetty} R., {Ostriker} E.~C., 2006, ApJ, 647, 997

\bibitem[{{Spitzer}(1942)}]{1942ApJ....95..329S}
{Spitzer}, L. J., 1942, ApJ, 95, 329

\bibitem[{{Tacconi} {et~al}\mbox{.}(2013){Tacconi}, {Neri}, {Genzel}, {Combes},
  {Bolatto}, {Cooper}, {Wuyts}, {Bournaud}, {Burkert}, {Comerford}, {Cox},
  {Davis}, {F{\"o}rster Schreiber}, {Garc{\'{\i}}a-Burillo}, {Gracia-Carpio},
  {Lutz}, {Naab}, {Newman}, {Omont}, {Saintonge}, {Shapiro Griffin}, {Shapley},
  {Sternberg}, \& {Weiner}}]{2013ApJ...768...74T}
{Tacconi} L.~J. {et~al.}, 2013, ApJ, 768, 74

\bibitem[{{Teyssier}(2002)}]{2002A&A...385..337T}
{Teyssier} R., 2002, A\&A, 385, 337

\bibitem[{{Toomre}(1964)}]{1964ApJ...139.1217T}
{Toomre} A., 1964, ApJ, 139, 1217

\bibitem[{{Truelove} {et~al}\mbox{.}(1997){Truelove}, {Klein}, {McKee},
  {Holliman}, {Howell}, \& {Greenough}}]{1997ApJ...489L.179T}
{Truelove} J.~K., {Klein} R.~I., {McKee} C.~F., {Holliman}, II J.~H., {Howell}
  L.~H., {Greenough} J.~A., 1997, ApJ, 489, L179

\bibitem[{{Wang} {et~al}\mbox{.}(2010){Wang}, {Klessen}, {Dullemond}, {van den
  Bosch}, \& {Fuchs}}]{Wang:2010wr}
{Wang} H.-H., {Klessen} R.~S., {Dullemond} C.~P., {van den Bosch} F.~C.,
  {Fuchs} B., 2010, MNRAS, 407, 705

\end{thebibliography}

\appendix
\section{Derivation of the reduction factor of the potential due to the disc thickness}
\label{App:Derivation of the reduction factor of the potential due to the disc thicknes}
Density perturbations correlate with variations in the gravitational field and are specified by the local gravitational potential $\Phi_{0}$, which is given by the solution of the Poisson equation in the stationary form with \citep[e.g.][]{Binney:2011vb, Wang:2010wr}
\begin{equation}
\label{eq:potentialperturbed}
\Phi_0(k,x,z)  = - \frac{2 \upi G  \Sigma}{\mid k \mid} \; \mathrm{e}^{\mathrm{i} k x - \mid k z\mid}. 
\end{equation}
Here, $\Sigma$ is the total surface density, $k$ the wavenumber, and $x=R-R_{0}$ the position near a given location $R_0$, and $G$ is the gravitational constant. $\Phi_0$ approaches zero for low surface densities $\Sigma$ and large distances $z$ from the plane. We consider only $k \geq 0$ and replace the wavenumber by the wavelength $\lambda = 2 \upi / k$, then
\begin{equation}
\label{eq:potentialperturbed_lambda} 
\Phi_0(\lambda,x,z)  = -  G \; \Sigma \; \lambda \; \mathrm{e}^{ \frac{2 \upi}{\lambda} ( \mathrm{i} x - \mid z\mid) }. 
\end{equation}
For the total potential of an axisymmetric three-dimensional disc with a finite thickness, we sum-up all contributions, generated from infinitesimally thin layers at all vertical distances $h$  (e.g. \citet{Wang:2010wr}) \\
\begin{equation}
\label{eq:phi_tot} 
\Phi_{\mathrm{tot}}(\lambda, x, z)  =  \int_{-\infty}^{\infty} \Phi_0(\lambda, x, z-h) \; t(h) \; dh.
\end{equation}
Here, $t(h)$ represents the vertical distribution of the gas density and satisfies the normalization condition
\begin{equation}
\label{eq:normalization}
\int_{-\infty}^{+\infty} t(h) \; dh = 1,
\end{equation}
while $\Sigma \: t(h) \: dh$ is the surface density of an infinitesimally thin layer that is located at $h$ above the mid-plane. Then
\begin{equation}
\label{eq:phi_tot_integral} 
\Phi_{\mathrm{tot}}(\lambda, x, z)  = -  G \; \Sigma \; \lambda \; \mathrm{e}^{\frac{2 \upi}{\lambda} \mathrm{i} x}   \int_{-\infty}^{\infty}  \mathrm{e}^{- \frac{2 \upi}{\lambda} \mid z - h \mid} \; t(h) \; dh,
\end{equation}
while the reduction factor is defined due to equation (\ref{eq:phi_tot_reduction_factor}) as
\begin{equation}
\label{eq:F_definition}
F(\lambda) =  \int_{-\infty}^{\infty}  \mathrm{e}^{- \frac{2 \upi}{\lambda} \mid h \mid} \; t(h) \; dh.
\end{equation}
Hence with the normalization equation (\ref{eq:normalization}) $t(h) = \sech^{2}(h/z_{0}) / (2 z_{0})$ and equation (\ref{eq:F_definition}) leads to equation (\ref{eq:integralsechhyperbolfunction}) and for the exponential profile to equation (\ref{eq:integralexponentialfunction}).

\section{Calculation of the fastest growing wavelength for the exponential profile approximation}
\label{App:Calculation of the fastest growing wavelength for the exp-profile approximation}
Here, we calculate the analytical solution of the fastest growing perturbation wavelength $\lambda_{f_{\mathrm{exp}}} = \lambda_{f_{\mathrm{exp}}}(\Sigma, c_{\mathrm{s}},z_{0})$, the global minimum of the dispersion relation for the exponential thickness approximation equation (\ref{eq:disp_relation_exp}) for $\lambda > 0$
\begin{equation}
\frac{\upartial \omega_{\mathrm{exp}}^{2}} {\upartial \lambda} = 4 \upi^{2} \left( \frac{G \; \Sigma}{ (\lambda + 2 \upi z_{0})^{2}} - \frac{2 c_{\mathrm{s}}^{2}}{\lambda^{3}}  \right)   = 0 ,
\end{equation}
with $\frac{\upartial^{2} \omega_{\mathrm{exp}}^{2}} {\upartial^{2} \lambda} < 0$.\\
We only consider the real solution of three (two are imaginary), and is given by
\begin{equation}
\lambda_{f_{\mathrm{exp}}} = \frac{2 c_{\mathrm{s}}^{2}}{3 G \Sigma} + \frac{4 c_{\mathrm{s}}^{2} }{9 G^2 \Sigma^2}  \frac{(c_{\mathrm{s}}^{2} + 6 \upi G \, \Sigma \, z_{0})}{T_{A}} +  T_{A},
\end{equation}
with the substitution
\begin{dmath}
T_{\mathrm{A}} = \left(\sqrt{ \frac{16 \upi^4 c_{\mathrm{s}}^{4} z_{0}^{4} }{G^2 \Sigma^2}  + \frac{64 \upi^3 c_{\mathrm{s}}^{3} z_{0}^{3} }{27 G^3 \Sigma^3} } +  \frac{8 c_{\mathrm{s}}^{6}}{27 G^3 \Sigma^3} + \\ + \frac{8 \upi c_{\mathrm{s}}^{4} z_0}{3 G^2 \Sigma^2} + \frac{4 \upi^2 c_{\mathrm{s}}^{2} z_{0}^{2}}{G \Sigma} \right)^{1/3} .
\end{dmath}
With the scaleheight $z_{0}$ equation (\ref{eq:scale-height_}) and the fastest growing wavelength $\lambda_{f_0}$ equation (\ref{eq:fastest_growing_wavelength_thin_disk}) of the thin disc approximation, which simplifies to $\lambda_{f_{\mathrm{exp}}}(\Sigma, c_{\mathrm{s}}, z_{0}) \rightarrow \lambda_{f_{\mathrm{exp}}}(\Sigma, c_{\mathrm{s}}) \rightarrow \lambda_{f_{\mathrm{exp}}}(  \lambda_{f_0})$, and is
\begin{equation}
\lambda_{f_{\mathrm{exp}}} =  \lambda_{f_0} \; \left( \frac{1}{3} + \frac{7}{9 T_{\mathrm{A}}} + T_{\mathrm{A}}   \right),   
\end{equation}
with
\begin{equation}
T_{\mathrm{A}} = \left(\sqrt{ \frac{1 }{4}  + \frac{1}{27} } + \frac{1}{27} + \frac{1}{3} + \frac{1}{2} \right)^{1/3} ,
\end{equation}
leading finally to
\begin{equation}
\lambda_{f_{\mathrm{exp}}} \simeq 2.148 \; \lambda_{f_0},   
\end{equation}
where we call the constant factor $A_{\mathrm{exp}} \simeq 2.148$.

\section{Hydrostatic equilibrium}
\label{App:Hydrostatic equilibrium} 
\begin{figure}
\includegraphics[width=84mm]{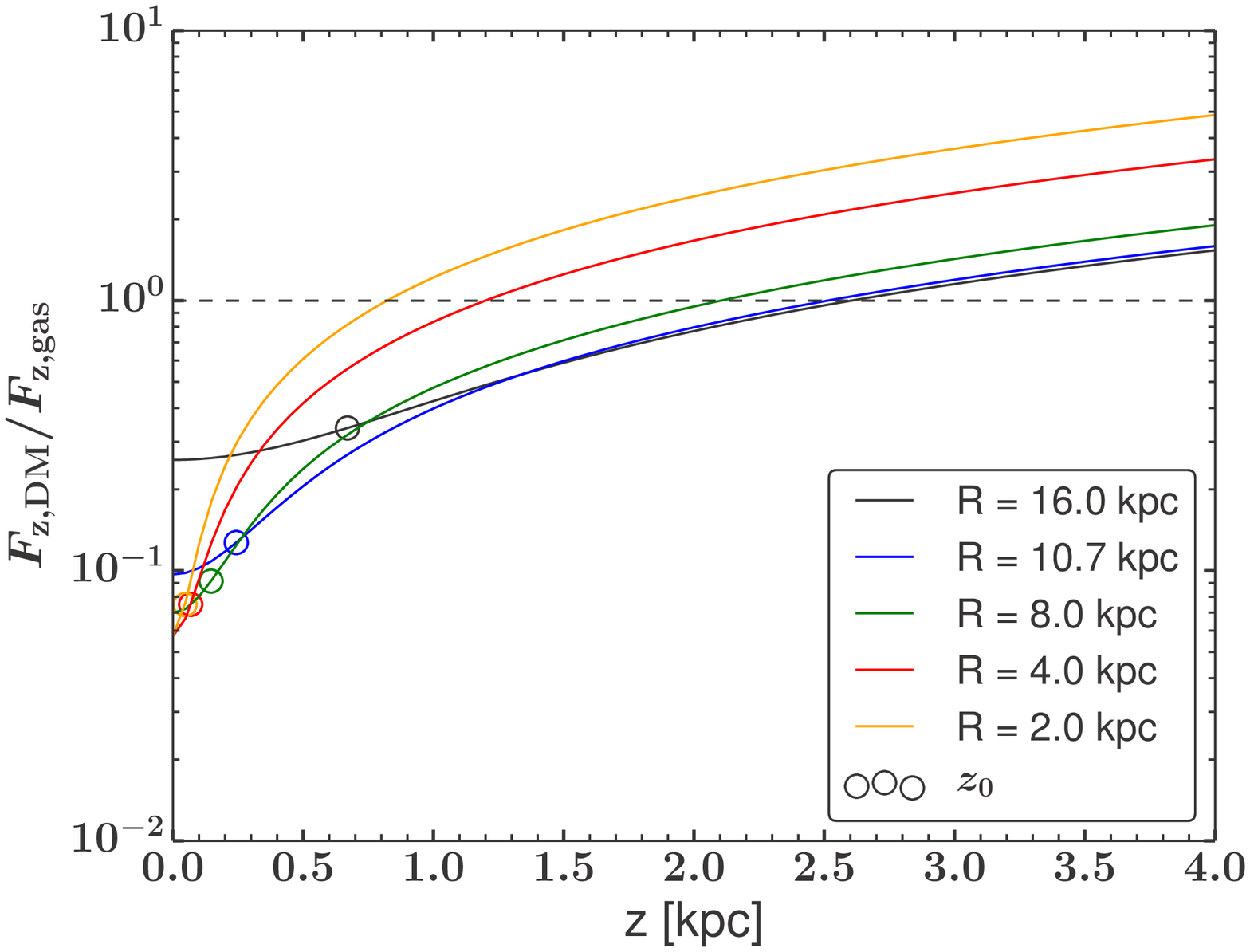} 
\caption{The ratio of the vertical forces corresponding to the gas disc and the dark matter halo for different disc radii. With distance from the mid-plane the dark matter plays an increasingly stronger role. The open circles mark the scaleheights $z_0$ at the corresponding radius $R$ and lie clearly in the regime, where the self-gravity of the disc still strongly dominates.
\label{fig:force_ratio_z}} 
\end{figure}

In this section, we describe how to bring our isothermal disc in hydrostatic equilibrium. We proceed with two requirements.\\
1. The disc has to have an exponential surface density
\begin{equation}
\label{App:first_requirenment}
\Sigma(R) = \Sigma_{0} \, \exp \left( - \frac{R}{h} \right),
\end{equation}
with the central surface density $\Sigma_{0}$ and the scalelength $h$.\\
2. A self-gravitating gas disc with its isothermal vertical structure in hydrostatic equilibrium is given by the $\sech^2$ density profile \citep{1942ApJ....95..329S}
\begin{equation}
\label{App:second_requirenment}
\rho_{\mathrm{gas}}(R,z) = \rho(R,0) \,  \sech^{2} \left( \frac{z}{z_{0}(R)} \right),
\end{equation}
where we keep the mid-plane density distribution $\rho(R,0)$ unknown and the scaleheight $z_{0}(R)$ a radial dependent quantity \citep{Wang:2010wr}.\\
The surface density is the integral over the vertical gas distribution (equation \ref{App:second_requirenment}) and gives
\begin{equation}
\Sigma(R) = \int_{-\infty}^{\infty} \rho_{\mathrm{gas}}(R,z) \, \dz = 2 \, \rho(R,0) \, z_{0}(R),
\end{equation}
and is therefore with our first requirement (equation \ref{App:first_requirenment})
\begin{equation}
\label{App:third_statement}
\Sigma(R) = 2 \, \rho(R,0) \, z_{0}(R) =  \Sigma_{0} \, \exp \left( - \frac{R}{h} \right),  
\end{equation}
where the scaleheight for the constant $c_{\mathrm{s}}$ is \citep{Wang:2010wr, Binney:2011vb}
\begin{equation}
\label{App:scale_height}
z_{0}(R) = \frac{c_{\mathrm{s}}}{\sqrt{2 \upi G \, \rho(R,0) }}.
\end{equation}
Equations \ref{App:third_statement} and \ref{App:scale_height} lead to
\begin{equation}
\rho_{\mathrm{gas}}(R,z) = \rho_{\mathrm{c}} \, \exp \left( - \frac{R}{h} \right)^{2} \,  \sech^{2} \left( \frac{z}{z_{0}(R)} \right),
\end{equation}
with the constant central density
\begin{equation}
 \rho_{\mathrm{c}} = \frac{\Sigma_{0}^2 \upi G}{2 c_{\mathrm{s}}^2}.
\end{equation}
The calculation does not take an external potential into account and, therefore, requires the vertical force of self-gravitating gas to dominate the external dark matter halo potential within the disc (see \citet{Wang:2010wr}):
\begin{equation}
F_{\mathrm{z,gas}} \gg F_{\mathrm{z,DM}},
\end{equation}
which is fulfilled for our disc setup (see Fig. \ref{fig:force_ratio_z}).

\label{lastpage}

\end{document}